\documentclass{article}

\usepackage{arxiv}

\usepackage[utf8]{inputenc} 
\usepackage[T1]{fontenc}    
\usepackage{hyperref}       
\usepackage{url}            
\usepackage{booktabs}       
\usepackage{amsfonts}       
\usepackage{nicefrac}       
\usepackage{microtype}      
\usepackage{lipsum}		
\usepackage{graphicx}
\usepackage{natbib}
\usepackage{doi}

\def\bSig\mathbf{\Sigma}

\usepackage{hyperref}
\usepackage{xr-hyper}
\usepackage{hyperref}
\usepackage[]{graphicx}
\usepackage[]{color}
\usepackage{caption}

\usepackage{alltt}

\usepackage{natbib}
\usepackage{bibentry}

\usepackage{amsthm}
\usepackage{amsmath}
\usepackage{amssymb}
\usepackage{mathabx}

\usepackage[section]{placeins}
\usepackage{float} 
\usepackage{subfig} 
\usepackage{amssymb}
\usepackage{setspace}
\usepackage{rotating}
\usepackage{graphicx}
\usepackage{booktabs} 
\usepackage{stmaryrd}
\usepackage{times}
\usepackage{bm}
\usepackage{indentfirst}

\usepackage{multirow}
\usepackage{multicol}
\usepackage{array}
\usepackage{booktabs}
\usepackage[table]{xcolor}

\usepackage{soul}

\usepackage[compact]{titlesec}
\titlespacing{\section}{0pt}{2ex}{1ex}
\titlespacing{\subsection}{0pt}{1ex}{0ex}
\titlespacing{\subsubsection}{0pt}{0.5ex}{0ex}

\title{Case Weighted Adaptive Power Priors for Hybrid Control Analyses with Time-to-Event Data}

\author{Evan Kwiatkowski$^{1,*}$, 
Jiawen Zhu$^{2}$,
Xiao Li$^{2}$,
Herbert Pang$^{2}$,
Grazyna Lieberman$^{2}$, and Matthew A. Psioda$^{3}$ \\
$^{1}$University of Texas MD Anderson Cancer Center, Department of Biostatistics, Houston, TX\\
$^{2}$Genentech, Department of Biostatistics, South San Francisco, CA\\
$^{3}$University of North Carolina, Department of Biostatistics,
Chapel Hill, NC \\
$^*$\texttt{ekwiatkowski@mdanderson.org}}

\renewcommand{\shorttitle}{CASE WEIGHTED POWER PRIORS FOR HYBRID
CONTROL ANALYSES}

\hypersetup{
pdftitle={A template for the arxiv style},
pdfsubject={q-bio.NC, q-bio.QM},
pdfauthor={David S.~Hippocampus, Elias D.~Striatum},
pdfkeywords={First keyword, Second keyword, More},
}

\begin{document}
\maketitle

\begin{abstract}
We develop a method for hybrid analyses that uses external controls to augment internal control arms in randomized controlled trials (RCT) where the degree of borrowing is determined based on similarity between RCT and external control patients to account for systematic differences (e.g. unmeasured confounders). The method represents a novel extension of the power prior where discounting weights are computed separately for each external control based on compatibility with the randomized control data. The discounting weights are determined using the predictive distribution for the external controls derived via the posterior distribution for time-to-event parameters estimated from the RCT. This method is applied using a proportional hazards regression model with piecewise constant baseline hazard. A simulation study and a real-data example are presented based on a completed trial in non-small cell lung cancer. It is shown that the case weighted adaptive power prior provides robust inference under various forms of incompatibility between the external controls and RCT population.
\end{abstract}


\section{Introduction}

RCT control data and real world data (RWD) are referred to as compatible if they are generated by statistical processes underpinned by similar parameters \citep{Brard2019}.
%
%
It is reasonable to expect there will be some degree of incompatibility between RCT controls and external controls obtained from RWD. 
For example, external controls may have systematically different unobserved characteristics compared with the RCT subjects or, if not concurrent, changes in medical practice may have occurred between the time associated with RWD data collection and the RCT.
Challenges to using RWD to construct hybrid control arms include systematic bias (e.g. drift in population baseline risk due to evolution of standard of care or some other unknown reason), data availability lag (i.e. some event times not available at the time of the RWD cut manifest as erroneous administrative censoring), and measurement error in key prognostic variables.
Regulatory authorities are typically concerned that the introduction of RWD (or more generally, external information) could inflate the trial's type I error rate or lead to bias in the estimated treatment effect.
It is generally understood that controlling type I error at a conventional level in cases where information borrowing is used generally limits or completely eliminates the benefits of borrowing \citep{Psioda2018b}.
If the control data are compatible then analyses using information borrowing can have improved operating characteristics, such as lower mean squared error and higher power.

\textcolor{black}{RWD are often used to construct an informative prior distribution for parameters common with the RCT data model, such as prognostic covariates separate from the treatment effect.}
An issue with using an informative prior is the possibility of incompatibility between the prior and the observed data, referred to as prior-data conflict. 
%
%
%
The prior predictive distribution informs which observable data are plausible based on the prior distribution and can be used to assess the compatibility of a prior and the data.
When RWD are surprising (i.e. unlikely in a probabilistic sense) based on a prior predictive distribution derived from the RCT data, this signals that something may be systematically different between the generative process for the RWD and the RCT data \citep{Lek2019}.
\cite{Box1980} describes how the prior predictive distribution can be used to assess the compatibility of a prior and the data (i.e. Box's $p$-value). 
\textcolor{black}{This can be used to identify unsuitable priors that conflict with the observed data \citep{Evans2006}, a concept that has been used recently in adaptive trial design (e.g. \cite{Psioda2020}).}

The power prior of \cite{Ibrahim2000} provides an ``equal but discounted" approach to managing the influence of external controls.
%
%
%
%
A prior is constructed by down-weighting the amount of borrowing from the external data using a weight that is fixed, treated as random (e.g. the normalized power prior \citep{Duan2006}), or dynamically estimated by techniques such as model selection criterion (e.g., penalized likelihood, marginal likelihood - see \cite{Ibrahim2015} Section 5 and the references therein) or empirical Bayes-type approaches based on marginal likelihood \citep{Gravestock2017}.
%
%
%
With a dynamically estimated weight, the degree of borrowing can depend on some measure of the compatibility between the RCT and external control data.
%
%

%

The commensurate prior \citep{Hobbs2011} provides robustness in incorporating historical information that is biased due to confounding which effects all the subjects equally (i.e. shifted baseline hazard). 
For this method applied to the problem at hand, the priors for the model parameters for the current data are centered at the corresponding parameters for the RWD, and a distribution is assumed for a commensurability parameter representing the difference in model parameters between the RCT data and the RWD. 
This method is similar to the meta-analytic predictive (MAP) approach \citep{Neuenschwander2010} in that these methods assume a variance parameter for the between-trial heterogeneity, but it is better suited to the case of a single external control dataset rather than multiple studies \citep{Dejardin2018}.
%
%


In this paper, we develop hybrid control analysis methods that use external controls to augment RCT control arms.
\textcolor{black}{We assess compatibility of each external control individually based on assessments of prior-data conflict using Box's $p$-value.}
This \textit{case-specific} weighting can capture incompatibility among a subset of the external controls and is ideally suited for scenarios where the RWD is \textit{partially} contaminated (i.e. an unobserved confounder affects a subset of the external controls).
%
%
%
We aim to have this method maintain traditional type I error control and to perform well in scenarios where the RWD is biased due to confounding which effects all the external controls equally. 
We use the commensurate prior for comparison 
since it explicitly includes a commensurability parameter that can represent drift in baseline hazard in the setting of survival analysis.
The motivating example for the simulation studies and real data analysis is a completed trial in non-small cell lung cancer for which there are relevant potential external controls, which is re-imagined as if a hybrid control arm were included as a part of the trial’s design.
%

\textcolor{black}{The rest of this paper is organized as follows: 
In Section \ref{sec:ch2methods}, we 
define a compatibility function for the external controls that is used to determine the case-specific weights and construct the case weighted power prior. 
%
%
In Section \ref{sec:ch2results}, we provide a simulation study demonstrating the case weighted power prior under different types of confounding in the external data. 
In Section \ref{sec:realdata} we provide an analysis using the case weighted power prior on a real data set.
%
%
We close the paper with some discussion in Section \ref{sec:ch2discussion}.}
\section{Methods}
\label{sec:ch2methods}
\subsection{External Likelihood}
For RCT subject $i$, $y_{1i}$ is the \text{observation time}, $\nu_{1i}$ is the \text{event indicator}, $\mathbf{x}_{1i}$ is the \text{covariate vector}, and $z_{1i}$ is the binary treatment indicator. 
The observation time $y_{1i}$ is given as $y_{1i}=\text{min}\{t_{1i},c_{1i}\}$, where $t_{1i}$ is the event time and $c_{1i}$ is the censoring time. 
Let $\nu_{1i}=I(y_{1i}\leq c_{1i})$ be the indicator that an event is observed. 
Denote the RCT data by $\mathbf{D}_1=\{(y_{1i},\nu_{1i},\mathbf{x}_{1i},z_{1i}):i=1,...n_1\}$, where $n_1$ are the number of RCT subjects. 
%

 

We consider a proportional hazards model with 
baseline hazard parameters $\boldsymbol{\lambda}$, covariate effect regression parameters $\boldsymbol{\beta}$, and treatment effect $\gamma$, with all unknown parameters denoted by $\boldsymbol{\theta}=\{\boldsymbol{\lambda},\boldsymbol{\beta},\gamma\}$. 
Let $\mathbf{x}_i$ and $z_{i}$ denote the covariate vector and binary treatment indicator for RCT subject $i$, respectively. The hazard for RCT subject $i$ is represented as
$
h_i(t|\boldsymbol{\theta})=h_0(t|\boldsymbol{\lambda})\text{exp}(\mathbf{x}_i'\boldsymbol{\beta}+z_i\gamma),
$
and the hazard for external control $j$ is represented as 
$
h_j(t|\boldsymbol{\lambda},\boldsymbol{\beta} )=h_0(t|\boldsymbol{\lambda})\text{exp}(\mathbf{x}_j'\boldsymbol{\beta}).
$
The same proportional hazards model (i.e., baseline hazard and covariate effect) for the outcome is used for the RCT and external control data to allow compatibility assessments to be made based on these shared parameters. 
Partition the time axis into $K$ intervals using $0=\tau_0<\tau_1<\tau_2<...<\tau_K=\infty$, and let $\boldsymbol{\tau}=(\tau_0,...,\tau_K)$. 
The partition of the time axis is chosen to capture important changes in the hazard rate while not introducing too many parameters. 
While there are formal regularization approaches for determining the partition (e.g. \cite{Bouaziz2016}), henceforth we will pre-specify $K$ and choose $\boldsymbol{\tau}$ to have an equal number of events in each interval.  
Let $\mathcal{I}_k=(\tau_{k-1},\tau_k]$. Let $\boldsymbol{\lambda}=(\lambda_1,...,\lambda_K)^T$. 
The baseline hazard $h_0(t \big| \boldsymbol{\lambda} )$ is taken as piecewise constant with $h_0(t \big| \boldsymbol{\lambda} ) = \lambda_k\text{ for }t \in \mathcal{I}_k$.

Denote the external control data by $\mathbf{D}_0=\{(y_{0j},\nu_{0j},\mathbf{x}_{0j}):j=1,...,n_0\}$, with all quantities analogous to the RCT data.
 Define the weighted likelihood for the external controls with both subject- and interval-specific weights by
\begin{align}\label{eq:external_likelihood_subject_interval}
    \prod_{j=1}^{n_0}\mathcal{L}(\boldsymbol{\beta},\lambda|\mathbf{D}_{0j}, \mathbf{a}_j)  =\prod_{j=1}^{n_0}\left\{(\lambda_{K_j}\text{exp}(x_j^T\boldsymbol{\beta}))^{a_{j,K_j}\nu_j}\prod_{k=1}^{K_j}\text{exp}\left\{ -a_{j,k}\lambda_k H_{j,k} \text{exp}(x_j^T\boldsymbol{\beta})\right\}\right\}_,
\end{align}
where $\mathbf{D}_{0j}=\{(y_{0j},\nu_{0j},\mathbf{x}_{0j})\}$ is the data for external control $j$, $\mathbf{a}_j=\{a_{j,1},...,a_{j,K_j}\}$ is a vector of interval-specific weights for external control $j$, and $K_j\in \{1,...,K\}$ is the index for the interval such that $y_j\in \mathcal{I}_{K_j}$, and $H_{j,k}$ represents at-risk time during interval $\mathcal{I}_{j}$ for external control $j$ for $k=1,...,K_j$.
\textcolor{black}{Throughout we assume that, conditional on covariates, the censorship times are independent of the event times for both the RCT and external data.}
\subsection{Case Weights}
To determine the value of weights $\mathbf{a}_j=\{a_{j,1},...,a_{j,K_j}\}$ for external control $j$, we assess the compatibility of the time at risk in interval $k$, $H_{j,k}$, relative to its predictive distribution derived from the RCT data to determine whether the value is extreme relative to what would be expected. 
The computation of the weights $\mathbf{a}_j$ is related to the memoryless property of the exponential distribution, which states that the probability a subject experiences an event after time $t$ (given they do not have an event before that point) \textcolor{black}{does not depend on} the probability that they experience an event prior to time $t$.
This property also applies to the piecewise exponential distribution and can be used for straightforward simulation of event times by simulating data for each time interval between cut-points in $\boldsymbol{\tau}$ from independent exponential distributions.
This affords us the ability to create a separate case weight for each interval for each external control by assessing the compatibility of the time at risk in the interval relative to its predictive distribution.

To compute this predictive distribution, it is necessary to specify and estimate a model for random censoring as it occurs in the RWD because the time at risk in an interval is a function of both the event and censorship distributions.
This model for censoring does not need to be specified for the RCT data (assuming event and censoring times are independent), since the compatibility of external control observation times will be assessed with respect to parameters in the distribution for event times. 
\textcolor{black}{The hazard for censorship for external control $j$ may be represented as
$
h^c_j(c|\boldsymbol{\lambda}^c,\boldsymbol{\beta}^c )=h_0(c|\boldsymbol{\lambda}^c)\text{exp}(\mathbf{x}_j'\boldsymbol{\beta}^c)$,
where $\boldsymbol{\lambda}^c=\{{\lambda}^c_{1},...,{\lambda}^c_{K}\}$ are the baseline hazard parameters, and $\boldsymbol{\beta}^c$ are covariate effect regression parameters. Without loss of generality, we consider the same intervals $\mathcal{I}_k$ that were used for the event distributions in the RCT and external control data.}

Let $y_{j}^{\text{rep}}$ be defined as replicated data \citep{Gelman2013} that could have been observed using the same model and value for $\theta$ that produced the randomized control data, the same censoring model that produced stochastic censoring in the external controls, and the same covariate vector $\mathbf{x}_j$ as external control $j$.
The predictive distribution for $y_{j}^{\text{rep}}$ is given by
\begin{align}\label{eq:pred}
p( y_{j}^{\text{rep}} \big| \mathbf{D}_{1}, \mathbf{D}_{0} )= 
\int p(y_{j}^{\text{rep}}  \big|\mathbf{x}_{j}, \boldsymbol{\lambda},\boldsymbol{\lambda}^c,\boldsymbol{\beta} )
\pi(\boldsymbol{\lambda},\boldsymbol{\beta} \big|\mathbf{D}_{1} ) \pi(\boldsymbol{\lambda}^c|\mathbf{D_0}) d\boldsymbol{\lambda} d\boldsymbol{\lambda}^c d\boldsymbol{\beta},
\end{align}
where $p(y_{j}^{\text{rep}}  \big|\boldsymbol{x}_{j}, \boldsymbol{\lambda},\boldsymbol{\lambda}^c,\boldsymbol{\beta} )$ is the density of the observation time for the $j$th external control,  $\pi(\boldsymbol{\lambda},\boldsymbol{\beta} \big|\mathbf{D}_{1} )$ is the posterior distribution 
for $\boldsymbol{\lambda}$ and $\boldsymbol{\beta}$ based on the randomized control data, and $\pi(\boldsymbol{\lambda}^c|\mathbf{D_0})$ is the posterior distribution 
for $\boldsymbol{\lambda}^c$ based on the RWD.
For a particular interval $\mathcal{I}_k$ (assuming external control $j$ is at risk in interval $k$), the predictive distribution from equation \eqref{eq:pred} becomes
\begin{align}\label{eq:pred_interval}
p( y_{j,k}^{\text{rep}} \big| \mathbf{D}_{1}, \mathbf{D}_{0} )= 
\int p(y_{j,k}^{\text{rep}}  \big|\mathbf{x}_{j}, \lambda_k, \lambda^c_k,\boldsymbol{\beta} )
\pi(\boldsymbol{\lambda},\boldsymbol{\beta} \big|\mathbf{D}_{1} ) \pi(\boldsymbol{\lambda}^c|\mathbf{D_0}) d\lambda_k d\lambda^c_k d\boldsymbol{\beta}.
\end{align}

\textcolor{black}{It is our objective to use the value of the predictive density from equation \eqref{eq:pred_interval} to assess compatibility of the observed RWD, such that observation times that are extreme relative to their predictive distribution will have comparatively lower predictive density values. Using a proportional hazards model with piecewise constant baseline hazards, it is necessary to transform the predictive distribution from equation \eqref{eq:pred_interval} so that the mode does not occur at time zero. Using the predictive density from equation \eqref{eq:pred_interval} would only allow observations that are higher than anticipated to be determined as incompatible based on their predictive density value.} This transformation will use a function $t$ such that 
$w_{j,k}^{\text{rep}} = t(y_{j,k}^{\text{rep}} \big| \mathbf{D}_{1}, \mathbf{D}_{0} )$
is approximately normally distributed (e.g. $t(x) = \log(x)$ is used henceforth), \textcolor{black}{allowing for observation times that are either lower or higher than anticipated to be evaluated as more extreme}. 
The weight $a_{j,k}$ is assigned as the probability of observing data as or more extreme (i.e. less likely) than the observed external control value $w_{j,k}=t(y_{j,k})$, and is an implementation of Box's $p$-value \citep{Box1980}. Formally, this is given by
\begin{align}\label{eq:ajk}
a_{j,k} = \text{Pr}\left[p( w_{j,k}^{\text{rep}}) \leq 
p( w_{j,k}) \right],
\end{align}
where the probability (i.e. expectation) is taken with respect to the density $p(w_{j,k}^{\text{rep}})$.

\textcolor{black}{For the interval $k=K_j$ containing the observation time, $y_{j,k}=H_{j,k}$ is the value that is transformed 
and that will be used in the compatibility assessment in equation \eqref{eq:ajk}. For $k<K_j$, we define the left-truncated random variable 
$y_{j,k}^*=(y_{j,k}|y_{j,k}>H_{j,k})$,
where $y_{j,k}^*$ is defined as the hypothetical observation time according to the hazards $h_j(t|\boldsymbol{\lambda},\boldsymbol{\beta} )=h_0(t|\boldsymbol{\lambda})\text{exp}(\mathbf{x}_j'\boldsymbol{\beta})$ and $h_j(c)=h_0(c|\boldsymbol{\lambda}^c)\text{exp}(\mathbf{x}_j'\boldsymbol{\beta}^c)$ for events and censoring, respectively. 
Then we compute $a_{j,k}$ as the expected Box's $p$-value from equation \eqref{eq:ajk} taken over the distribution of the hypothetical observation times $y_{j,k}^*$, which is given by 
\begin{align}\label{eq:ajk_star}
a_{j,k} = E_{y_{j,k}^*}\left[\text{Pr}\left(p({ w_{j,k}^{\text{rep}}}^*) \leq p( w_{j,k}^*) \right) \right],
\end{align}
where ${w_{j,k}^{\text{rep}}}^*$ and $w_{j,k}^*$ are based on the transformed values $y_{j,k}^*$.}
%
If there is perfect compatibility of the RCT controls and external controls, then the weights $a_{j,k}$ will be uniformly distributed since the shared parameters are equivalent and the posterior predictive distribution is continuous \citep{Gelman2013}.
Further details on the computational implementation are in Web Appendix A. 

\subsection{Case Weighted Power Priors}
We use the weights $a_{j,k}$ defined using equations \eqref{eq:ajk} and \eqref{eq:ajk_star} to create a case weighted power prior as a generalization of the fixed-weight power prior $\pi_0(\boldsymbol{\theta}|\mathbf{D}_0,a_0)\propto [\mathcal{L}(\boldsymbol{\theta}| \mathbf{D}_0)]^{a_0} \pi_0(\boldsymbol{\theta}).$ 
We replace $[\mathcal{L}(\boldsymbol{\theta}| \mathbf{D}_0)]^{a_0}$ in the fixed-weight power prior with the weighted likelihood for the external controls with both subject- and interval-specific weights from equation \eqref{eq:external_likelihood_subject_interval} with the addition of a calibration function which influences the operating characteristics of the analysis. 
This calibration function $ h(a_{j,k},\overline{A})$ is applied to each weight $a_{j,k}$ individually, and also is based on the average case weight for all external controls $\overline{A}={\sum_{j=1}^{n_0}\sum_{k=1}^{K_j}a_{j,k}}/{\sum_{j=1}^{n_0} K_j}$. 
The calibrated weighted likelihood becomes
\begin{align}\label{eq:f2-external-likeihood}
    \prod_{j=1}^{n_0}\mathcal{L}(\boldsymbol{\beta},\lambda|&\mathbf{D}_{0j}, h(\mathbf{a}_j,\overline{A})) \nonumber \\ &=\prod_{j=1}^{n_0}\left\{(\lambda_{K_j}\text{exp}(x_j^T\boldsymbol{\beta}))^{h(a_{j,K_j},\overline{A})\nu_j} \prod_{k=1}^{K_j}\text{exp}\left\{ -h(a_{j,k},\overline{A})\lambda_k H_{j,k} \text{exp}(x_j^T\boldsymbol{\beta})\right\}\right\}_.
\end{align}

%
%
The resulting power prior using the likelihood in equation \eqref{eq:f2-external-likeihood} using $h(a_{j,k},\overline{A})=f_p(a_{j,k})$ defines the case weighted power prior, where the function $f_p$ is referred to as the case weight shrinkage function.
The parameter $p$ controls the degree to which the case weights are tempered towards the constant 0.5 (their expected value under perfect compatibility) to counterbalance the modest type I error rate inflation which would arise from using $a_{j,k}$ directly in equation \eqref{eq:f2-external-likeihood} (i.e. using untransformed case weights), with the conservative type I error rate of the fixed weight power prior in order to produce an analysis with controlled type I error rate at the nominal level. See Web Appendix B and C for additional explanation.

The resulting power prior using the the likelihood in equation \eqref{eq:f2-external-likeihood} using $h(a_{j,k},\overline{A})=f_p(a_{j,k})g_c(\overline{A})$ defines the \textit{discounted} case weighted power prior, where the function $g_c$ is referred to as the uniform discounting function.
The function $g_c$ is based on a predetermined level of maximum tolerated type I error rate (e.g., $0.15$ is used henceforth) in the event that there is a shift in baseline hazard for all external controls. A similar process was implemented by \cite{Psioda2018a}, and also could be framed as a predetermined maximum level of power reduction for a shift in baseline hazard for all external controls.
The parameter $c$ controls the degree to which all case weights are reduced in value, which would tend towards a no-borrowing design, based on the difference of $\overline{A}$ with 0.5 (its expected value under perfect compatibility). 
%
%
%

%
\textcolor{black}{The calibration procedure considers the given model (e.g. proportional hazards model with specific set of covariates) and given sample sizes for RWD and RCT data under the assumption of compatible external controls, and therefore is separate from the actual RWD outcomes (see details in Web Appendix D).}
%
%
%
%
%
%
The calibrated weights are also used to derive the case weighted commensurate prior which serves as a comparison method (see details in Web Appendix E).

\section{Simulation Studies}\label{sec:ch2results}




\subsection{Simulation Setup}\label{sec:ch2simsetup}
As a motivating example, we consider NCT02008227 (OAK study)
\cite{Rittmeyer2017}, a global, multicenter, open-label, randomized, controlled study which evaluated the efficacy and safety of atezolizumab compared with docetaxel in participants with locally advanced or metastatic non-small cell lung cancer (NSCLC) after failure with platinum-containing chemotherapy.  
Among 850 participants randomized 1:1, an analysis of overall survival using Cox partial likelihood yields an estimated hazard ratio of $0.73$ with $95\%$ CI $(0.62-0.86)$ in favor of atezolizumab. 

We consider RWD from the nationwide (EHR)-derived longitudinal Flatiron Health database, comprised of de-identified patient-level structured and unstructured data curated via technology-enabled abstraction originating from $\sim$280 US cancer clinics ($\sim$800 sites of care) \citep{MA2020, Birnbaum2020}. 
Existing research has used Flatiron Health databases for external control analyses in oncology studies \citep{Ventz2019, Lewis2019, Schmidli2019}. 
We consider $526$ external controls that meet OAK inclusion/exclusion criteria, henceforth referred to as NSCLC RWD. 
%
%

Compatibility is assessed based on models which adjust for covariate effects; therefore, the methods considered take into account differences in measured characteristics. 
In fact, the data sources have different distributions of observed covariates (i.e., external controls average age 67.1 vs. RCT 63.2, external controls male 55.6\% vs. RCT 66.2\%), and the covariate effects are independently estimated from the fitted model parameters from the respective data source and therefore are distinct.
We adjust for sex and age as measured covariates thought to be of prognostic value, and introduce confounding through a covariate (i.e. unobserved confounder) which represents systematic differences in the two hazards that are not explainable by measured covariates. 
We considered a hazard model for RCT subject $i$ given by
$h_i(t|\boldsymbol{\theta})=h_0(t|\boldsymbol{\lambda})\text{exp}(\text{age}_i\beta_{1}+I(\text{sex = male})_i\beta_2+z_i\gamma)$,
and let the hazard for external control $j$ be
$h_j(t|\boldsymbol{\lambda},\boldsymbol{\beta})=h_0(t|\boldsymbol{\lambda})\text{exp}(\text{age}_j\beta_{1}+I(\text{sex = male})_j\beta_2+ x_{3j}\beta_3)$, where $x_{3j}$ is the confounding covariate for external control $j$.

We consider three types of confounding based on the confounding covariate $x_{3j}$. ``Partial contamination" occurs when the unobserved confounder affects a subset of the external controls, indicating non-exchangeability of a latent subpopulation which could arise from a data quality issue (e.g., site-specific measurement error) in a clinical trial or incorrect information in an external control’s electronic health record which provides crucial data in a RWD cohort. 
This type of measurement error is likely common, but few methods are available to explicitly account for it. 
In particular, $x_3=\text{log}(2^m)$, with $\text{Pr}(m=0)=0.68$ indicating no confounding and $\text{Pr}(m=2k)=\text{Pr}(m=-2k)=0.02$ for $k=1,...,8$ indicating varying degrees of confounding among the subpopulation.
``Shift confounding" represents a shift in the baseline hazard for all external controls. In particular, $x_3=1$. ``Partial shift confounding" represents a shift in baseline hazard for the latter interval of survival time (i.e. $t > \tau_1$). In particular, $x_3=I[t > \tau_1]$ \textcolor{black}{is the indicator that $t$ falls in the second of the two intervals which will be used in this illustration.}


The values of $\beta_3$ used in the covariate effect $x_{3j}\beta_3$ range from $-\text{log}(3)$ to $\text{log}(3)$ representing hazard rations between -3 and 3. The baseline hazard $h_0\left(t \big| \lambda \right)$ and the censoring distribution represented by the hazard $h_0(c|\boldsymbol{\lambda}^c)$ are taken as piecewise constant with
$0=\tau_0<\tau_1<\tau_2 = \infty$ and $h_0\left(t \big| \boldsymbol{\lambda}^c \right) = \lambda^c_k\text{ for }t \in \left( \tau_{k-1} ,\tau_{k} \right]$. \textcolor{black}{For the purposes of this illustration, the cutpoints used to generate the data were assumed known, and the same cutpoints were used in the analysis.}

We consider an analysis which incorporates data from {200} subsampled RCT treated subjects, {100} subsampled RCT controls, and {100} subsampled external controls in an augmented analysis under different assumptions for the covariate $x_3$ and the magnitude of $\beta_3$. For the simulation studies, we find estimated values of ${\boldsymbol{\lambda}}$, $\boldsymbol{\lambda}^c$, $\gamma$, $\beta_1$, and $\beta_2$ from an analysis of the RCT subjects, using cutpoints $\boldsymbol{\tau}$ chosen to have an equal number of events in each interval. 

We consider two scenarios for the prevalence of censoring in the external control population, both of which are based on modifications of the fitted value $\boldsymbol{\lambda}^c$ from the actual RCT data. ``Low Censoring" will consider the baseline hazard for censoring to be $1.4\cdot \boldsymbol{\lambda}^c$, and ``High Censoring" will consider $0.9\cdot\boldsymbol{\lambda}^c$.
%
%
Note that the censoring distribution in the RCT and external controls are not assumed to be equivalent; the RCT data are used to provide an initial estimate $\boldsymbol{\lambda}^c$ which is further perturbed by the multiplicative factors of $\{0.9, 1.4\}$. Since the number of external controls are fixed, the ``Low Censoring" scenario is associated with more events and thus more information contained in the external control data, and the ``High Censoring" scenario is associated with fewer events and less information.

The hypothesis under consideration is the one-sided hypothesis $H_0:\gamma\geq 0$ vs. $H_1:\gamma<0$. 
This hypothesis was evaluated by computing the posterior probability of $\gamma<0$ being less than 0.025. \textcolor{black}{The true parameter values considered are $\gamma=0$ (for the null hypothesis) and $\gamma=\text{log}(0.73)$ (for the alternative hypothesis).}
The simulation study summarizes $10,000$ repetitions per value of $\beta_3$ used to produce confounding. All simulations were performed using R version 4.1.2 \citep{R2017}.
%
 %
%
Analyses using the commensurate prior are fit using the Hamiltonian Monte Carlo algorithm using STAN version 2.27.0 and \texttt{cmdstanr} version 0.4.0.

\subsection{Distribution of Case Weights}

\begin{figure}[!htbp]
\begin{center}
\includegraphics[width=6in]{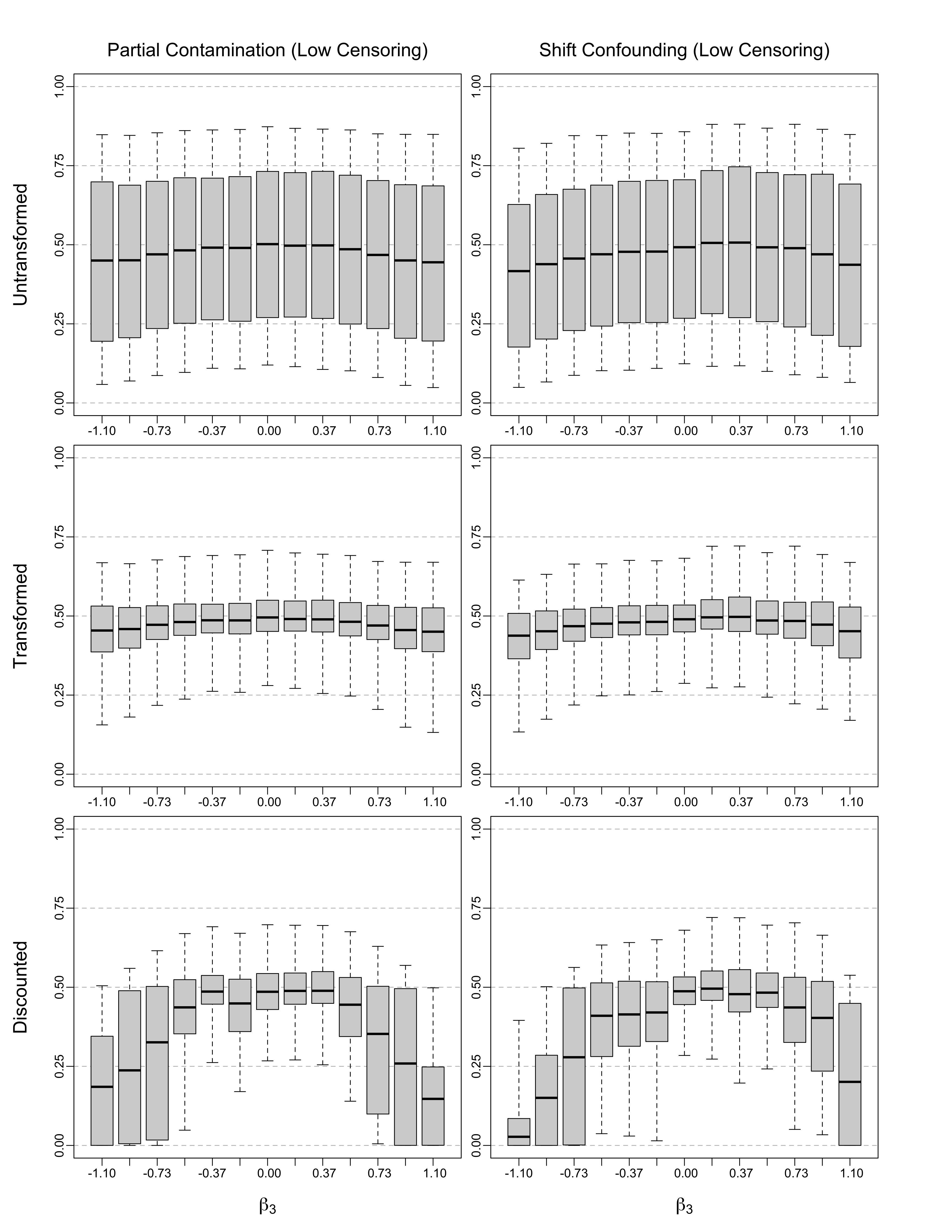}
\caption{
Case weight distributions by data generation scenario and analysis method. Untransformed = untransformed case weighted power prior, Transformed = case weighted power prior; Discounted = discounted case weighted power prior with maximum type I error rate under shift confounding calibrated at 0.15.
Boxplot with mean, inter-quartile range, and 10th/90th percentiles of case weights.}
\label{fig:ch2weights}
\end{center}
\end{figure}

%
Figure \ref{fig:ch2weights} shows the distribution of the case weights for the partial contamination and shift confounding scenarios averaged across simulated trials. 
For the untransformed weights, when $\beta_3=0$, the external controls are exchangeable with the RCT controls and the weights $a_{j,k}$ are uniformly distributed on the unit interval with a mean value of $0.5$. 
As the magnitude of $\beta_3$ increases, the distribution of the case weights begins to differ from a uniform distribution. 

The case weights transformed by the case weight shrinkage function $f_p(a_{j,k})$ are similar to the untransformed weights in their average, however there is less overall dispersion. 
Notice that while the magnitude of $\beta_3$ increases, the average case weight decreases, and the range of the 75th to the 90th percentile lengthens considerably. \textcolor{black}{This is because the effect of the confounding on survival time increases as the magnitude of $\beta_3$ increases, causing lower case weights to be assigned to those observations most impacted. The transformed weights $f_p(a_{j,k})$ approach zero as $a_{j,k}$ approaches zero, so those observations most impacted by the confounding could be assigned an arbitrarily low case weight.} It is this shrinking of the case weights around $0.5$ that enables the case weighted power prior to have a controlled type I error rate; the transformed weights are closer to the fixed value of $0.5$ which mimics a power prior with a fixed weight and a conservative type I error rate. 

The case weights transformed by both the case weight shrinkage function and the uniform discounting function $f_p(a_{j,k})g_c(\overline{A})$ are similar to the transformed case weights $f_p(a_{j,k})$ for values of $|\beta_3|$ near zero, and are nearly equivalent when $\beta_3= 0$. This is because there is little dataset-level incompatibility detected in the external controls. However, as $\beta_3$ increases, the average case weight drops substantially, as the difference between $\overline{A}$ from $0.5$ increases. It is this drop in case weights that enables the discounted case weighted power prior to have a calibrated maximum type I error rate under shift confounding; for large levels of incompatibility in the external data, the amount of borrowing decreases substantially.

Table \ref{tbl:weights} shows the average untransformed case weights by amount of censoring and confounding. Under the partial contamination scenario, the average case weights fall in the intervals $(0.468, 0.500)$ and $(0.447, 0.500)$ for the high and low censoring scenarios, respectively. This interval is narrower in the high censoring scenario because the parameters for the external control population are estimated with less precision with fewer events among the external controls, making it harder to discern incompatibility with the RCT data. This same pattern also is seen with shift and partial shift confounding.
Table \ref{tbl:weights} shows that the average case weights for events and censored observations generally are similar, which highlights that compatibility is based on the observation time which can be either an event or an instance of censoring.
For the cases of partial contamination and shift confounding, the untransformed case weights corresponding to survival time in the first and second intervals are generally similar because the confounding equally affects the hazards in both intervals. However, for the case of the partial shift confounding, the case weights in the first interval of survival time are uniformly distributed with average values of 0.5 for both low and high censoring. This is because there is no incompatibility in the external controls among the first segment of survival time; only the second interval has the effect of the confounder. 

\begin{table}[!htbp]
\centering
\caption{Average case weights by confounding scenario and amount of censoring, averaged first within a dataset and then across simulated datasets. $a_0$ = average case weights overall; $a_{0,\text{evt}}$ = average case weights for events; $a_{0, \text{cen}}$ = average case weights for censored observations, $a_{0, \text{int 1}}$ = average case weights over $(\tau_0,\tau_1]$, $a_{0, \text{int 2}}$ = average compatiblity weights over $(\tau_1,\tau_2]$.}
\begin{tabular}{cccccccccc}
  \hline
  \begin{tabular}{c}Censoring \\Amount\end{tabular}   & \begin{tabular}{c}Type of \\Confounding\end{tabular}   & $\beta_3$ & \begin{tabular}{c}Average \\ \# Events\end{tabular}   & 
  $a_0$ & $a_{0,\text{evt}}$ & $a_{0, \text{cen}}$ & 
  $a_{0, \text{int 1}}$ & $a_{0, \text{int 2}}$ \\
  \hline
 &  & -1.099 & 94.916 & 0.447 & 0.450 & 0.437 & 0.465 & 0.415 \\ 
   &  & -0.549 & 96.772 & 0.481 & 0.487 & 0.462 & 0.498 & 0.452 \\ 
Low   & Partial & 0.000 & 97.122 & 0.498 & 0.510 & 0.464 & 0.504 & 0.488 \\ 
   &  Contamination& 0.549 & 96.671 & 0.482 & 0.488 & 0.464 & 0.498 & 0.453 \\ 
   &  & 1.099 & 94.614 & 0.447 & 0.450 & 0.438 & 0.465 & 0.416 \\ 
   \hline
    &  & -1.099 & 92.886 & 0.408 & 0.408 & 0.408 & 0.441 & 0.368 \\ 
   &  & -0.549 & 95.690 & 0.463 & 0.477 & 0.433 & 0.471 & 0.453 \\ 
Low   & Shift & 0.000 & 97.120 & 0.499 & 0.510 & 0.465 & 0.504 & 0.488 \\ 
   &  & 0.549 & 97.369 & 0.502 & 0.497 & 0.524 & 0.513 & 0.474 \\ 
   &  & 1.099 & 97.025 & 0.446 & 0.425 & 0.607 & 0.447 & 0.444 \\ 
   \hline
    &  & -1.099 & 94.626 & 0.452 & 0.452 & 0.454 & 0.504 & 0.367 \\ 
   &  & -0.549 & 96.281 & 0.484 & 0.491 & 0.463 & 0.504 & 0.451 \\ 
Low   & Partial Shift  & 0.000 & 97.105 & 0.499 & 0.510 & 0.466 & 0.505 & 0.487 \\ 
   &  & 0.549 & 97.375 & 0.489 & 0.496 & 0.468 & 0.504 & 0.465 \\ 
   &  & 1.099 & 97.386 & 0.463 & 0.459 & 0.474 & 0.504 & 0.394 \\ 
   \hline
    &  & -1.099 & 54.096 & 0.468 & 0.461 & 0.474 & 0.463 & 0.480 \\ 
   &  & -0.549 & 54.595 & 0.490 & 0.501 & 0.480 & 0.491 & 0.486 \\ 
High   & Partial &  0.000 & 54.898 & 0.497 & 0.512 & 0.483 & 0.496 & 0.498 \\ 
   & Contamination & 0.549 & 54.406 & 0.490 & 0.500 & 0.481 & 0.491 & 0.487 \\ 
   &  & 1.099 & 53.985 & 0.468 & 0.459 & 0.475 & 0.461 & 0.483 \\ 
   \hline
    &  & -1.099 & 28.909 & 0.460 & 0.462 & 0.460 & 0.456 & 0.467 \\ 
   &  & -0.549 & 41.309 & 0.479 & 0.491 & 0.472 & 0.475 & 0.486 \\ 
High   & Shift & 0.000 & 54.904 & 0.497 & 0.512 & 0.483 & 0.496 & 0.498 \\ 
   &  & 0.549 & 67.722 & 0.498 & 0.506 & 0.486 & 0.502 & 0.482 \\ 
   &  & 1.099 & 78.405 & 0.456 & 0.454 & 0.465 & 0.461 & 0.426 \\ 
  \hline

 &  & -1.099 & 44.072 & 0.488 & 0.508 & 0.475 & 0.496 & 0.467 \\ 
   &  & -0.549 & 49.225 & 0.493 & 0.510 & 0.481 & 0.496 & 0.486 \\ 
High   & Partial Shift & 0.000 & 54.857 & 0.497 & 0.512 & 0.483 & 0.496 & 0.498 \\ 
   &  & 0.549 & 60.259 & 0.492 & 0.504 & 0.479 & 0.496 & 0.482 \\ 
   &  & 1.099 & 64.727 & 0.475 & 0.479 & 0.471 & 0.496 & 0.424 \\ 
   \hline
\end{tabular}
\label{tbl:weights}
\end{table}



\subsection{Operating Characteristics}
\subsubsection{Partial Contamination}\label{sec:ch2partialcontamination}
\begin{figure}[!htbp]
\begin{center}
\includegraphics[width=6.25in]{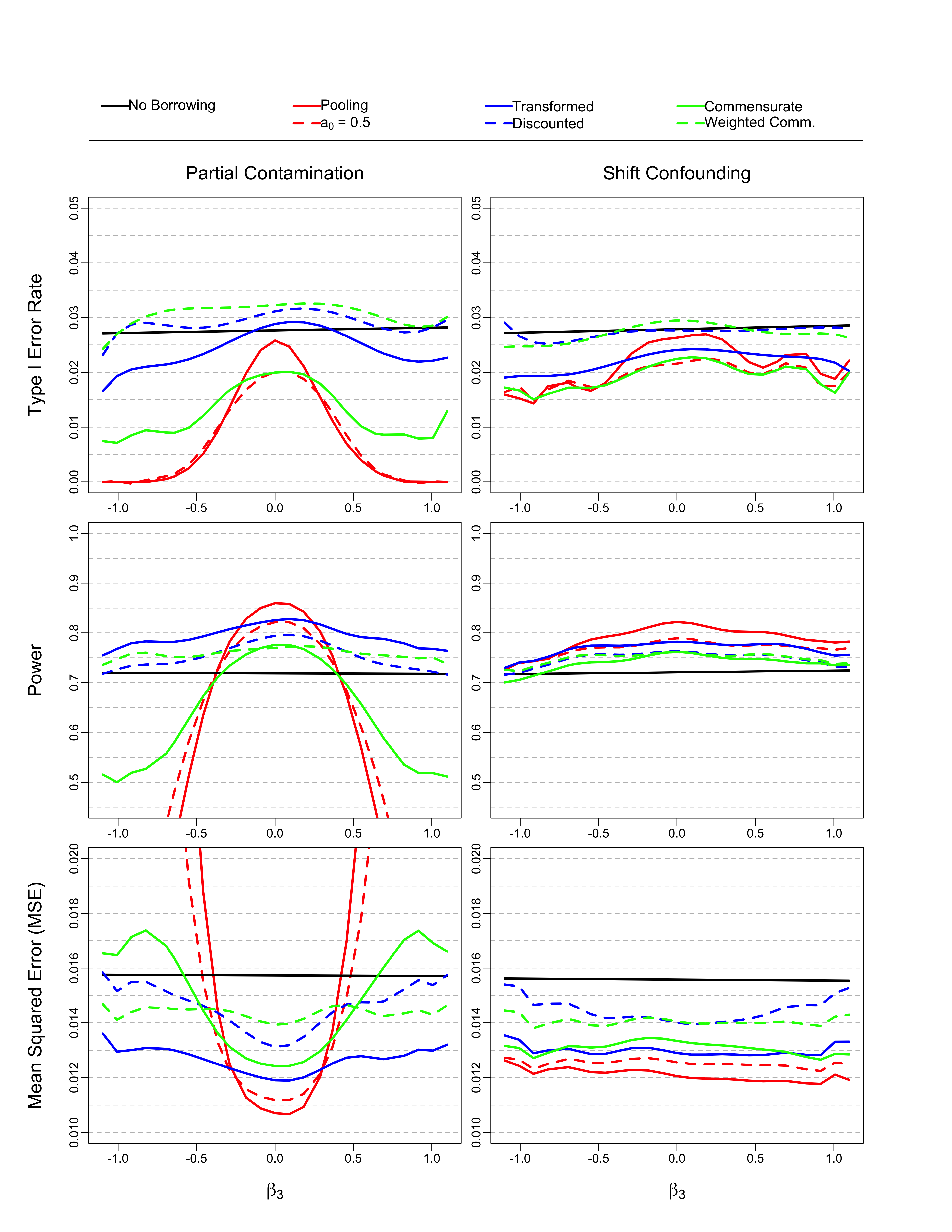}
\caption{Partial contamination scenario; selected operating characteristics by analysis method. Transformed = case weighted power prior; Discounted = discounted case weighted power prior with maximum type I error rate under shift confounding calibrated at 0.15.}
\label{fig:6panel-outlier}
\end{center}
\end{figure}

Figure \ref{fig:6panel-outlier} shows the operating characteristics of type I error, power, and mean squared error for the case weighted power prior for the partial contamination scenario. 
%
%
The power at the null value of $\beta_3=0$ is the highest for full-borrowing, followed by the fixed weight power prior with $a_0=0.5$. This is to be expected since those cases incorporate compatible external controls with a large weight. However, all fixed weight power priors suffer precipitous drops in power as as $|\beta_3|$ increases, since incompatible external control information is incorporated to a fixed (i.e. static) degree. The case weighted power priors maintain a high level of power for all values of $\beta_3$ considered since external controls with observed incompatibility are dynamically down-weighted. 
All the fixed weight power priors explored are shown to have relatively lower MSE compared to the case weighted approaches for an interval of $\beta_3$ around zero, then sharply increase to levels higher than no borrowing as $\beta_3$ increases. Case weighting maintains relatively low MSE for all values of $\beta_3$ considered.

The case weighted commensurate prior is shown to have a slightly inflated type I error rate when $\beta_3=0$. All weighted methods
are shown to have robustness to partial contamination, and maintain a high level of power for all $\beta_3$ values considered. The commensurate prior has the greatest reduction in power as $|\beta_3|$ increases, since the commensurate prior has no mechanism to dynamically weight outlying observations. \textcolor{black}{Note that the maximum type I error rate of the discounted case weighted power prior is less than the calibration value of $0.15$, since this calibration value was chosen under the assumption of a shift confounder affecting all external controls.}

\subsubsection{Shift Confounding}\label{sec:ch2shiftresults}
\begin{figure}[!htbp]
\begin{center}
\includegraphics[width=6.25in]{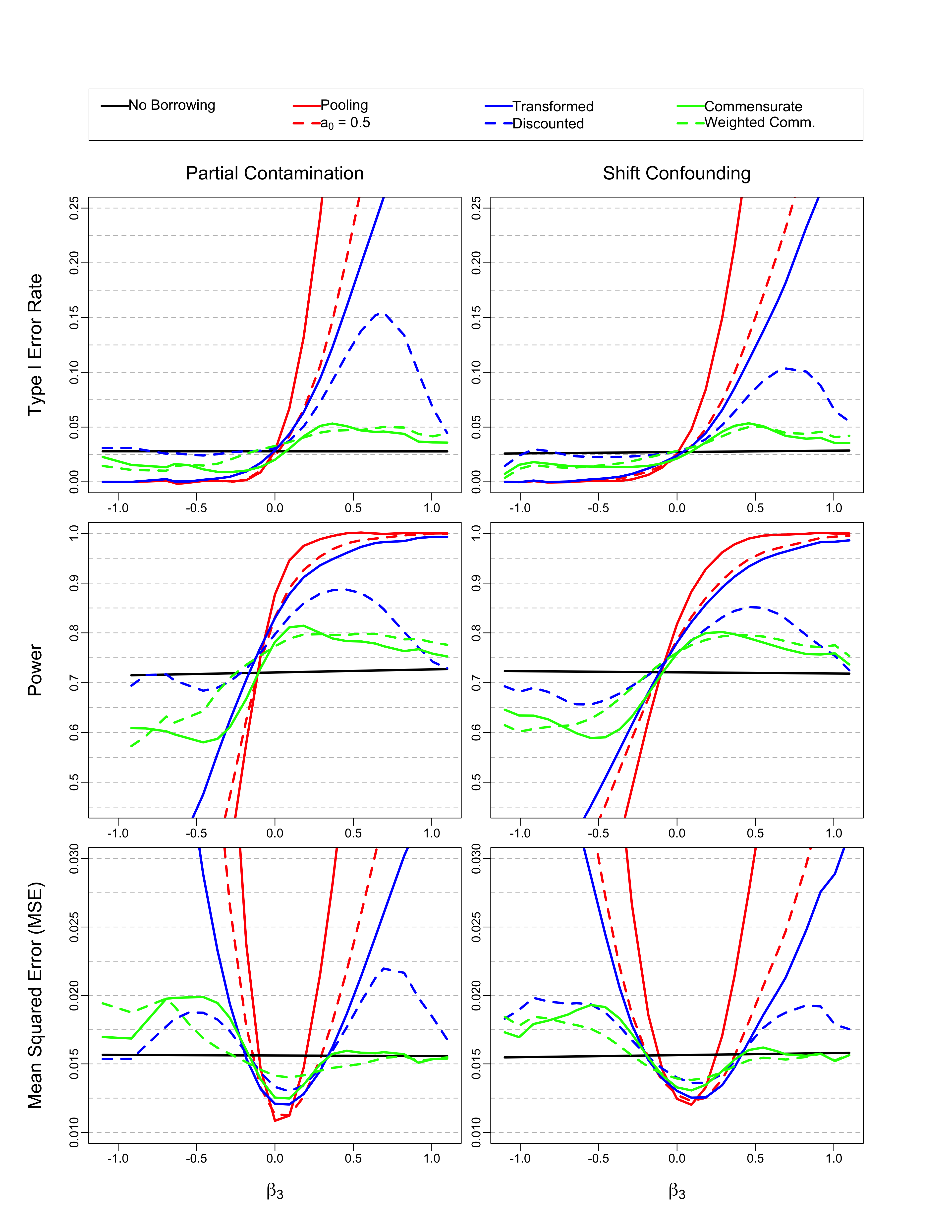}
\caption{Shift confounding scenario; selected operating characteristics by analysis method. Transformed = case weighted power prior; Discounted = discounted case weighted power prior with maximum type I error rate under shift confounding calibrated at 0.15.}
\label{fig:6panel-shift}
\end{center}
\end{figure}

Figure \ref{fig:6panel-shift} shows the operating characteristics of the case weighted power prior for the shift confounding scenario. 
%
Case weighting maintains higher power than fixed weight power priors for $\beta_3<0$ by dynamically down-weighting the external controls with observation times that are observed to be incompatible with the RCT data.
All fixed weight power priors are shown to have very high power for $\beta_3>0$, since \textcolor{black}{this results in an downward bias in the estimated hazards for controls resulting in a upward bias in the estimated treatment effect. The case weighted power prior has lower power in this case due to the down-weighting of the incompatible external controls, demonstrating that the case weighted power prior does not uniformly increase power in all scenarios.} As in the case of partial contamination, all fixed weight power priors have relatively low MSE for an interval of $\beta_3$ around zero, while the case weighted power prior has relatively low MSE for all values of $\beta_3$ considered.

The commensurate prior and case weighted commensurate prior behave similarly in terms of the type I error rate and power, while the commensurate prior has relatively lower MSE as $|\beta_3|$ increases. The commensurate priors with the chosen specification of hyperprior on the variance for the drift parameter have less spread around no borrowing for type I error and power. 
The maximum type I error rate for the commensurate prior methods under shift confounding is shown to be about $0.05$, which is much less than the discounted case weighted power prior with maximum type I error rate of $0.15$. 
\textcolor{black}{Note that the maximum type I error rate of the discounted case weighted power prior is less than the calibration value of $0.15$ for the high censoring scenario, since this calibration value was chosen under the low censoring mechanism.}

\subsubsection{Partial Shift Confounding}
The operating characteristics from the partial shift confounding scenario are generally similar to those of the shift confounding scenario displayed in Figure \ref{fig:6panel-shift} (see Web Appendix F). The magnitude of the type I error rate and power differences from the no borrowing case are shown to be less than those from the shift confounding scenario for all the power priors. This is to be expected since the partial shift confounding scenario corresponds to a lesser degree of incompatibility than the shift confounding scenario. 

\subsection{Choosing Number of Segments for Baseline Hazard}
It is necessary to give thoughtful consideration to the number of segments used for the baseline hazard in the analysis model. %
Web Table 5 shows that the lowest BIC occurs when the number of segments for the baseline hazards in the generating model matches the number of segments used in the analysis model. 
Consequently, these situations produce the more accurate results for the treatment effect estimation.
For example, when $\beta_3=0$ (i.e. compatible external controls) and $K_G=3$ segments are used in the generating hazard, the closest estimated hazard ratio to the generating value of $0.73$ is observed when $K_M=3$ segments are used in the analysis model.

\subsubsection{Vary Proportion of External Controls}
We explore the impact of modifying the number of external controls from 100 to either 200 or 50 while keeping the same amount of 200 subjects randomized to treatment and 100 randomized controls.
Modifying the number of external controls fundamentally alters the study’s operating characteristics (see Web Appendix F). For example, the power in the homogeneous case with 100 external controls is 0.86, which increases to 0.92 with 200 external controls and decreases to 0.82 with 50 external controls. Relatedly, the type I error is much more sensitive to confounding with more external controls, and less sensitive to confounding with fewer external controls. These are unavoidable consequences of having varying amounts of information informing the control group rather than distinctive shortcomings of the proposed methods.

Although the operating characteristics of the designs are altered, the proposed methods maintain their comparative advantages to the benchmark methods of no borrowing and pooling. When the number of external controls is modified, the type I error rate still is preserved. For the partial contamination scenario, the proposed methods still maintain higher power and lower MSE across the magnitude of the confounding variable $\beta_3$. For the shift confounding scenario, the proposed methods are more robust to the confounding than the pooling method, and provide increases in power and decreases in MSE relative to no borrowing when the unobserved confounding is minimal.

\subsubsection{Study Impact of Model Misspecification}
To study the impact of model misspecification, we consider delayed separation situations where the treatment effect is only present after a delay of 50 or 100 days. These modifications alter the operating characteristics of the study: for example, power is decreased from 0.86 to 0.82 with a 50-day delay and 0.77 with a 100-day delay (see Web Appendix F). Still, the proposed methods maintain their comparative advantages to the benchmark methods of no borrowing and pooling, such as higher power and lower MSE in the partial contamination scenario, and better robustness to shift confounding than pooling with increases in power and decreases in MSE relative to no borrowing when the unobserved confounding is minimal.
\section{Real Data Example}\label{sec:realdata}

\begin{figure}[!htbp]
\begin{center}
\includegraphics[width=6in]{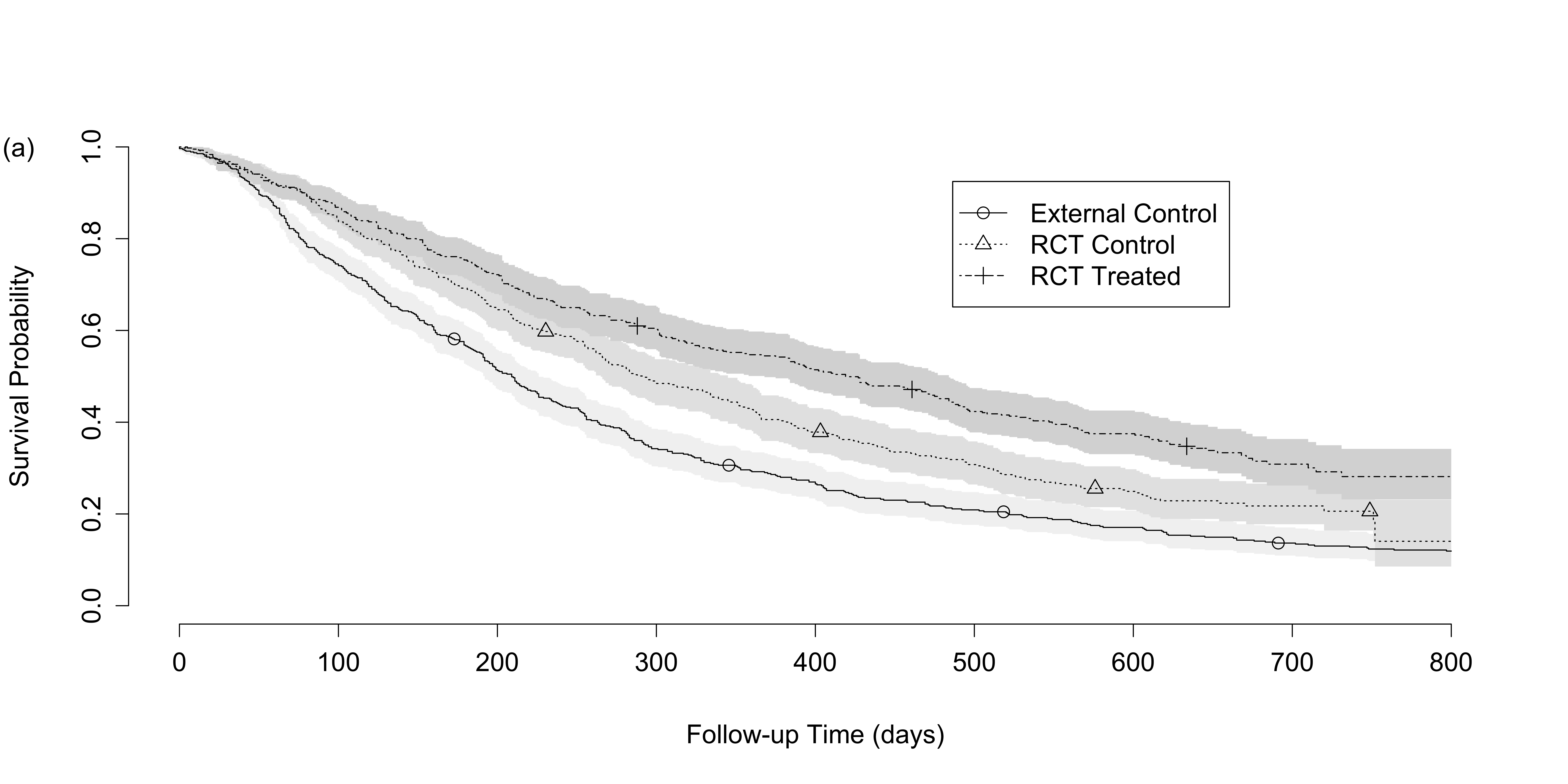}
\includegraphics[width=7in]{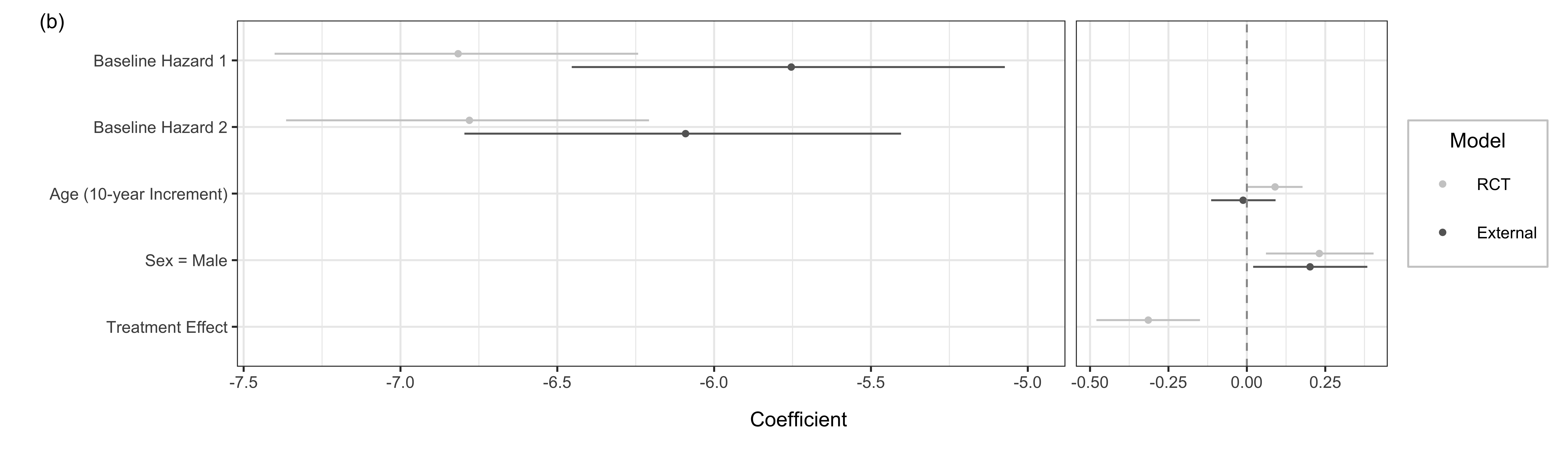}
\includegraphics[width=6.5in]{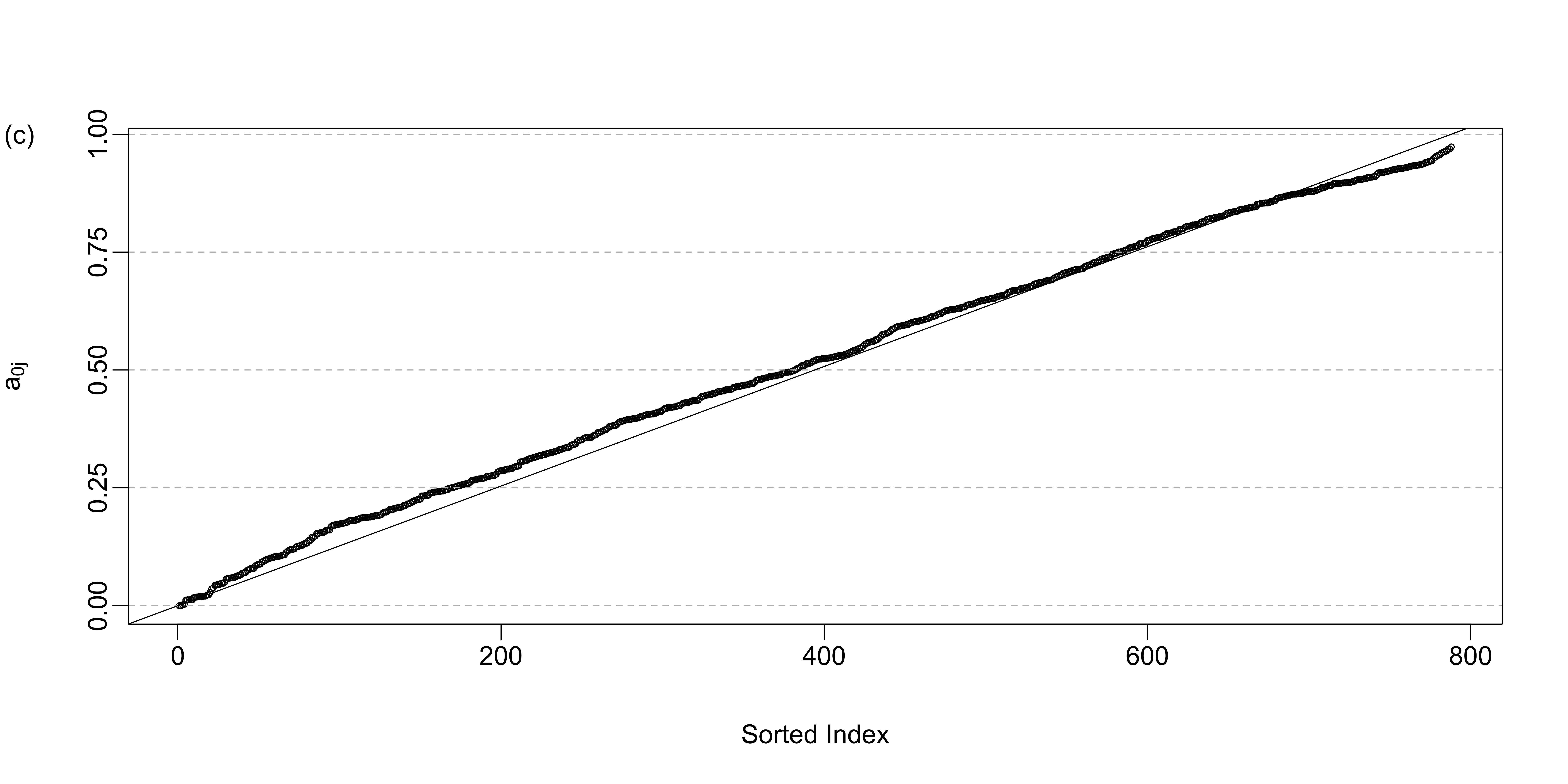}
\caption{(a) Kaplan-Meier curves for RCT and external data: (b) Fitted model coefficients for RCT and external data: (c) Compatibility weights for NSCLC RWD.}
\label{fig:realdata}
\end{center}
\end{figure}

We consider testing the adaptive borrowing method using all subjects from the real datasets which served as motivation for the simulation studies (i.e., 850 randomized subjects and 526 external controls). 
%
Figure 4(a) shows observed differences across the RCT and external datasets. The Kaplan-Meier curves show that the treatment arm has improved survival relative to the randomized control arm, which in turn has improved survival relative to the external control arm. Figure 4(b) shows fitted model coefficients for the RCT and external data, which implies compatibility in that the covariate effects of age and sex are highly similar between the datasets, and also implies incompatibility in that the baseline hazard components are lower for the RCT indicating improved survival (although the estimated cofficients have overlapping 95\% confidence intervals). An examination of the compatibility weights for the external controls in Figure \ref{fig:realdata}(c) demonstrates limited deviation from the anticipated uniform distribution. 
Table \ref{tbl:realdata} shows the estimate for the treatment effect and model fit diagnostics using the case weighted power prior. It is shown that when $\boldsymbol{\tau}$ has $3$ cutpoints, the BIC is the lowest, and the estimated hazard ratio associated with the treatment effect is $0.650$. This estimated hazard ratio is equivalent to the estimated hazard ratio for the pooling method (i.e., combining RCT and external data), although the pooling method has a slightly narrower credible interval. 
\begin{table}[p]
\centering
\caption{Treatment effect estimation and model fit diagnostics for OAK and NSCLC RWD datasets. $K_M$: number of baseline hazard segments used in the analysis model, $\overline{a_0}$: average case weight, HR: hazard ratio, CI W.: credible interval width, BIC: Bayesian Information Criterion.}
\begin{tabular}{ccccccccccc}
  \hline
& &
\multicolumn{3}{c}{Adaptive} & 
\multicolumn{3}{c}{No Borrowing} &
\multicolumn{3}{c}{Pooling}\\
                            \cline{3-11}
$K_M$	&	$\overline{a_0}$	&	HR	&	CI W.	&	BIC	&	HR	&	CI W.	&	BIC	&	HR	&	CI W. &	BIC	\\	
   \hline
1	&	0.474	&	0.670	&	0.197	&	14922.8	&	0.732	&	0.242	&	14931.7	&	0.653	&	0.182 &	14921.8	\\	
2	&	0.536	&	0.662	&	0.194	&	14940.1	&	0.730	&	0.242	&	14933.0	&	0.657	&	0.183 &	14956.6	\\	
3	&	0.544	&	0.650	&	0.190	&	14916.1	&	0.728	&	0.241	&	14935.7	&	0.650	&	0.181 &	14911.9	\\	
4	&	0.553	&	0.652	&	0.190	&	14975.2	&	0.729	&	0.242	&	14982.5	&	0.649	&	0.181 &	14992.7	\\	
5	&	0.556	&	0.651	&	0.190	&	14918.4	&	0.727	&	0.241	&	14942.7	&	0.643	&	0.180 &	14913.3	\\	
   \hline
\end{tabular}
\label{tbl:realdata}
\end{table}

\section{Discussion}\label{sec:ch2discussion}
The case weighted power prior provides a novel strategy for the incorporation of RWD into an analysis of RCT data.
The case weights provide a framework for comparatively more robust estimation of effects of interest compared to fixed weight power priors in scenarios where there is systematic incompatibility between RCT controls and external controls due to unmeasured confounding.
Using predictive distributions (e.g. Box's $p$-value) provides an intuitive metric of compatibility for comparing external control data to RCT data.
%
%
This method increases power in the case of compatibility but also succeeds in reducing the influence of incompatible external controls and limiting the increase in the type I error rate. 
Since each external control subject is assigned their own compatibility weights, this method can be directly applied to incorporating different sources of external controls into a single hybrid analysis.

A common model for the observation times is assumed for both the RCT and external control data. This approach is binding, but essential. If we are to evaluate how well RCT data predicts the external control data, we must have a common model to translate between them. Future research could extend the method to assume only a subset of the parameters are shared (e.g., only the baseline hazard parameters in the outcome model and not the covariates), and to use the external control data to estimate the covariate effects (as we have done with the censoring model).

\textcolor{black}{Addressing compatibility for time-to-event outcomes is an involved process, and we are aware of no other methods which can account for a partial shift in baseline hazard among external controls. A straightforward application of this methodology would be towards non-survival outcomes (e.g. normally distributed study endpoints). The capacity to detect prior-data conflict among external controls is heavily dependent on the total amount of information in the external controls through the number of events, which needs to be taken into consideration if this methodology is to be applied to study designs with interim analyses where fewer events are available among the external controls. The application of this method for trial design contexts is an area of future research.}


%

\textcolor{black}{Future work could involve testing robustness of the case weighted power priors under additional types of confounding between the RWD and RCT data, such as combinations of partial contamination and shifts in baseline hazards.} Data quality issues relating to RWD remain a persistent challenge, including difficulty in defining time zero for an external control which could result in immortal time bias favoring the RCT group \citep{Burcu2020}.

While there is interest in obtaining drug approvals using single-arm studies, our method determines case weights for the external controls by assessing how well the predictive distribution (based on the RCT) for their observation times data aligns with the actual observed data for the external controls, which is the uniqueness of hybrid control trials and analyses. In order for the proposed approach to be feasible, one must have some controls in the prospective trial. Thus, as constructed, this method would not be directly applicable to purely externally controlled trials.

\section*{Acknowledgements}

This work was partially supported by FDA grant NIH U01 FD007206 (Lieberman, Pang, Li, Zhu). EK was partially supported by the NCI T32 training
grant [5T32CA096520-15].

\bibliography{FinalDissertationReferences.bib}





\section*{Supporting Information} \label{s:Software}
Web Appendices, Tables, and Figures  referenced in Sections 2 and 3 are available with this paper at the Biometrics website on Wiley Online Library. A GitHub repository (https://github.com/kwiatkowski-evan/case-weighted-power-priors) contains the programs and other resources 
needed to reproduce the analyses in this paper. The method will be included in a future version of the \texttt{psborrow} package (https://cran.r-project.org/web/packages/psborrow/index.html).


\section*{Web Appendix A: Computational Details for Case Weighted Power Prior}
\subsubsection*{Power Prior }
\hspace{0.1in}The fixed-weight formulation of the power prior \citep{Ibrahim2000} is
\begin{align}\label{eq:power_prior}
\pi_0(\boldsymbol{\theta}|\mathbf{D}_0,a_0)\propto [\mathcal{L}(\boldsymbol{\theta}| \mathbf{D}_0)]^{a_0} \pi_0(\boldsymbol{\theta}).
\end{align}
We consider incorporating \textit{case-specific} weight parameters $A=\{\mathbf{a}_1,...,\mathbf{a}_{n_0}\}$, which modifies equation \eqref{eq:power_prior} to
\begin{align}\label{eq:power_prior_ajk}
\pi_0 (\boldsymbol{\lambda}, \boldsymbol{\beta}, \gamma \big| \mathbf{D}_0, A ) &\propto 
   { \left[\prod_{j=1}^{n_0} \mathcal{L}(\boldsymbol{\lambda}, \boldsymbol{\beta} \big|\mathbf{D}_{0j},\mathbf{a}_j)\right]}\pi_0( \boldsymbol{\lambda},\boldsymbol{\beta}, \gamma),
\end{align}
where $\pi_0( \boldsymbol{\lambda},\boldsymbol{\beta},\gamma)$ is an initial (typically noninformative) prior.
Inference is based on the posterior distribution
\begin{align}\label{eq:posterior_all}
    \pi(\boldsymbol{\beta},\boldsymbol{\lambda},\gamma|\mathbf{D}_1,\mathbf{D}_0)\propto \mathcal{L}(\boldsymbol{\beta},\boldsymbol{\lambda},\gamma|\mathbf{D}_1) \left[\prod_{j=1}^{n_0} \mathcal{L}(\boldsymbol{\lambda}, \boldsymbol{\beta} \big|\mathbf{D}_{0j},\mathbf{a}_j)\right] \pi_0(\boldsymbol{\beta},\boldsymbol{\lambda},\gamma).
\end{align}

\subsubsection*{Posterior Inference}
\hspace{0.1in}We use the Laplace approximation (i.e. multivariate normal approximation) to the posterior distribution from equation \eqref{eq:posterior_all} for the hazard ratio regression parameters, which eliminates the need for MCMC sampling. Let $\boldsymbol{\alpha}=\log(\boldsymbol{\lambda})$. The posterior distribution for $(\boldsymbol{\alpha},\boldsymbol{\beta},\gamma)$ converges in probability to the multivariate normal distribution
\begin{align}\label{eq:asymptotically-normal}
    (\boldsymbol{\alpha},\boldsymbol{\beta},\gamma|\mathbf{D}_1,\mathbf{D}_0,A)\sim \text{Normal}\left(\hat{\boldsymbol{\theta}}=\begin{bmatrix}
\hat{\boldsymbol{\alpha}}\\
\hat{\boldsymbol{\beta}}\\
\hat{\gamma}
\end{bmatrix},\hat{\Sigma}=
\begin{bmatrix}
\hat{\Sigma}_{11} & \hat{\Sigma}_{12} & \hat{\sigma}_{13} \\
\hat{\Sigma}_{21} & \hat{\Sigma}_{22} & \hat{\sigma}_{23} \\
\hat{\sigma}_{31} & \hat{\sigma}_{32} & \hat{\sigma}_{33} \\
\end{bmatrix}
\right)_,
\end{align}
where $\hat{\boldsymbol{\theta}}$ is the posterior mode, and $\hat{\Sigma}$ is the negative inverse of the Hessian matrix for the logarithm of the posterior evaluated at $\hat{\boldsymbol{\theta}}$. Matrix components of $\hat{\Sigma}$ are denoted with uppercase (e.g., $\hat{\Sigma}_{11}$) and vector components with lowercase (e.g., $\hat{\sigma}_{33}$). We make inference on $\gamma$ by considering the marginal distribution 
\begin{align}\label{eq:treatment-benefit}
   ( \gamma|\mathbf{D}_1,\mathbf{D}_0,a_0 )\sim \text{Normal} (\hat{\gamma}, \hat{\sigma}_{33}).
\end{align}
The approximation of the one-dimensional marginal distribution of interest in equation \eqref{eq:treatment-benefit} has been shown to be accurate even if it is based on full Laplacian approximation in equation \eqref{eq:asymptotically-normal} which may be less accurate for some components (e.g. the baseline hazard parameters; see \cite{Psioda2018}).


Let $\pi_0(\boldsymbol{\alpha},\boldsymbol{\beta},\gamma)\propto 1$ be a uniform improper prior, which will be used henceforth. Define $\delta_{i,k}=I(y_i\in \mathcal{I}_k)$. Then the posterior distribution from equation \eqref{eq:posterior_all} be expressed as
\begin{align*}
    \pi(\boldsymbol{\beta},\boldsymbol{\lambda},\gamma&|\mathbf{D}_1,\mathbf{D}_0)\propto \mathcal{L}(\boldsymbol{\beta},\boldsymbol{\lambda},\gamma|\mathbf{D}_1) \left[\prod_{j=1}^{n_0} \mathcal{L}(\boldsymbol{\lambda}, \boldsymbol{\beta} \big|\mathbf{D}_{0j},\mathbf{a}_j)\right] 
    \\\propto& \prod_{i=1}^{n_1} \prod_{k=1}^{K_i}(\lambda_{K_i}H_{i,k}\text{exp}(x_i^T\boldsymbol{\beta}+z_i\gamma))^{\delta_{i,k}\nu_i}\text{exp}\left\{-\lambda_k H_{i,k} \text{exp}(x_i^T\boldsymbol{\beta}+z_i\gamma)\right\}/(\delta_{i,k}\nu_i)\nonumber
    \\&\times\prod_{j=1}^{n_0} \prod_{k=1}^{K_j}(\lambda_{K_j}H_{j,k}\text{exp}(x_j^T\boldsymbol{\beta}))^{\delta_{j,k}a_{k,j}\nu_j}\text{exp}\left\{ -a_{j,k}\lambda_k H_{j,k} \text{exp}(x_j^T\boldsymbol{\beta})\right\}/(\delta_{j,k}a_{j,k}\nu_j)_,
\end{align*}
which is proportional to a product of subject- and interval-specific Poisson likelihoods.  Therefore, posterior inference can be assessed using software for Poisson regression (e.g. using the Newton-Raphson algorithm \citep{Stokes2000}) using a \text{log}-link for the rate parameter of the Poisson distribution, where $\text{log}(H_{i,k})$ (and $\text{log}(H_{j,k})$) is an offset and $a_{j,k}$ is a weight for the log-likelihood for interval $k$ of external control $j$. 
Other posterior distributions are found similarly, including $\pi(\boldsymbol{\lambda},\boldsymbol{\beta} \big|\mathbf{D}_{1} )$ and $\pi(\boldsymbol{\lambda}^c|\mathbf{D_0})$ which are used in the predictive distribution in Main Text equation (3).
\subsubsection*{Computing Case Weights}
\hspace{0.1in}Several computational approximations are used to determine the subject- and interval-specific case weights $a_{j,k}$ from Main Text equations (4-5). 
In order to generate $N$ samples (e.g. $N=10,000$) from $p( y_{j,k}^{\text{rep}} \big| \mathbf{D}_{1}, \mathbf{D}_{0} )$, the predictive distribution given in Main Text equation (3), we do the following: 
\begin{enumerate}
    \item Generate $N$ samples from each of the asymptotic normal approximations in equation \eqref{eq:asymptotically-normal} of the posteriors $\pi(\boldsymbol{\lambda},\boldsymbol{\beta} \big|\mathbf{D}_{1} )$ and $\pi(\boldsymbol{\lambda}^c|\mathbf{D_0})$.
    \item Use these quantities in Main Text equation (3) using the known form of the density $p(y_{j,k}^{\text{rep}}  \big|\mathbf{x}_{j}, \boldsymbol{\lambda},\boldsymbol{\lambda}^c,\boldsymbol{\beta} )$.
\end{enumerate}

\noindent In order to compute the case weights in Main Text equation (4), we do the following: 
\begin{enumerate}
    \item Transform the $N$ samples from the predictive distribution $p( y_{j,k}^{\text{rep}} \big| \mathbf{D}_{1}, \mathbf{D}_{0} )$ to get $N$ samples from $w_{j,k}^{\text{rep}} = t(y_{j,k}^{\text{rep}} \big| \mathbf{D}_{1}, \mathbf{D}_{0} )$.
    \item Compute the empirical density function of the $N$ samples $w_{j,k}^{\text{rep}}$ to get $p(w_{j,k}^{\text{rep}}\big| \mathbf{D}_{1}, \mathbf{D}_{0})$ (e.g. using kernel density estimation).
    \item Compute $a_{j,k}$ in Main Text equation (4) as the frequency that the empirical density evaluated at the $N$ samples are less than the empirical density evaluated at $w_{j,k}$.
\end{enumerate}

\noindent In order to compute $a_{j,k}$ in Main Text equation (5) when $k<K_j$, we do the following: 

\begin{enumerate}
    \item 
Consider sampling values from $y_{j,k}^*$ by generating imputed outcomes $y_{j,k}^{(n)}$, where $y_{j,k}^{(n)} = H_{j,k}+\text{min} \{ t_{j,k}^{(n)},c_{j,k}^{(n)}\}$, where $t_{j,k}^{(n)}$ and $c_{j,k}^{(n)}$ are generated according to the hazards $h_j(t|\boldsymbol{\lambda},\boldsymbol{\beta} )=h_0(t|\boldsymbol{\lambda})\text{exp}(\mathbf{x}_j'\boldsymbol{\beta})$ and $h_j(c)=h_0(c|\boldsymbol{\lambda}^c)\text{exp}(\mathbf{x}_j'\boldsymbol{\beta}^c)$ for events and censoring respectively for a fixed number of repetitions (e.g. $n=1,...,20$). %
    \item Then compute $a_{j,k}^{(n)}$ using the transformed values $w_{j,k}^{(n)}$, and set $a_{j,k}$ as the average of $\{a_{j,k}^{(n)}\}$.
\end{enumerate}

\section*{Web Appendix B: Inflated Type I Error Rate for Adaptive Case Weighted Power Prior}\label{sec:appendixb}
Let $D_1=\{y_{1i}\}_{i=1}^{n_1}$ and $D_0=\{y_{0j}\}_{j=1}^{n_0}$ represent sets of normally distributed random variables, each with unknown mean $\mu=0$ and known variance $\sigma^2=2$. 
Consider an initial uniform improper prior on $\mu$ and a fixed $0<a_0<1$ which will be used to weight $D_0$ in a power prior. Let $a_0=\pi_{1/2}$ correspond to power prior weights for $D_0$ that equal one with probability $0.5$ and zero otherwise, where the weights are randomly determined \textit{independent} of the outcomes. These weights provide a useful comparison since they will result in average posterior standard deviations for estimates of $\mu$ that are equal to those of adaptive weighting and $a_0=0.5$.
Consider the one-sided hypothesis $H_0: \mu\leq 0$. 
\begin{table}[ht]
\centering
\caption{Probably of rejecting $H_0$ and posterior summary statistics by power prior weighting method. Rej = probability of rejecting $H_0$; $P(\mu  > 0)$ = average posterior probability that $\mu > 0$; $\hat{\mu}$ = average posterior mean; $\hat{\sigma}$ = average posterior standard deviation.}
\begin{tabular}{lcccccccccc}

  \hline
  & & \multicolumn{3}{c}{Overall} & \multicolumn{3}{c}{Reject} & \multicolumn{3}{c}{Fail to Reject}\\
  \cline{3-11}Weight& Rej & $P(\mu > 0)$ & $\hat{\mu}$ & $\hat{\sigma}$& $P(\mu > 0)$ & $\hat{\mu}$  & $\hat{\sigma}$& $P(\mu > 0)$ & $\hat{\mu}$  & $\hat{\sigma}$\\
  \hline
$a_0=0$&0.025 & 0.499 & 0.000 & 0.141 & 0.987 & 0.330 & 0.141 & 0.487 & -0.009 & 0.141 \\ 
Adaptive&    0.036 & 0.499 & 0.000 & 0.115 & 0.989 & 0.277 & 0.116 & 0.481 & -0.011 & 0.115 \\ 
$a_0=\pi_{1/2}$ & 0.025 & 0.499 & 0.000 & 0.115 & 0.988 & 0.270 & 0.115 & 0.487 & -0.007 & 0.115 \\ 
$a_0=0.5$ & 0.016 & 0.499 & 0.000 & 0.115 & 0.986 & 0.264 & 0.115 & 0.491 & -0.005 & 0.115 \\ 
 $a_0=1$&  0.025 & 0.499 & 0.000 & 0.100 & 0.988 & 0.234 & 0.100 & 0.487 & -0.006 & 0.100 \\ 
   \hline
\end{tabular}
\end{table}

Note that the nominal type I error rate of $0.025$ is observed for the fixed weight power prior with weights of $a_0\in\{0,1\}$, and for $a_0=\pi_{1/2}$. The type I error rate for the adaptive case weighted power prior is $0.036$, and the type I error rate for the fixed weight power prior with $a_0=0.5$ is $0.016$. Interestingly, the average posterior standard deviation for $a_0=0.5$, $a_0=\pi_{1/2}$, and adaptive weighting are all $0.115$. Consider the average posterior mean of $\mu$ among cases that reject the null hypothesis. The average posterior mean is greatest for adaptive weighting at $0.277$, and smallest for $a_0=0.5$ at $0.264$, and has the intermediates value of $0.270$ when $a_0=\pi_{1/2}$. This shows that the use of adaptive case weights leads to erroneously rejecting $H_0$ by creating weighted estimates of $\mu$ that are larger than estimates produced by analysis methods with a controlled or a conservative type I error rate.

\section*{Web Appendix C: Conservative Type I Error Rate in Fixed Borrowing}\label{sec:appendix3}
We demonstrate the conservative type I error rate when borrowing a fixed amount of compatible information in the case of a normally distributed endpoint. Let $D_1=\{y_{1i}\}_{i=1}^{n_1}$ and $D_0=\{y_{0j}\}_{j=1}^{n_0}$ represent sets of normally distributed random variables, each with unknown mean $\mu$ and known variance $\sigma^2$. 
Consider an initial uniform improper prior on $\mu$ and a fixed $0<a_0<1$ which will be used to weight $D_0$ in a power prior. 
The posterior distribution of $\mu$ is 
\begin{align*}
    \pi(\mu|D_1,D_0,a_0)&\propto \pi(D_1|\mu)\pi(D_0|\mu)^{a_0}\\
    &\propto\exp \left\{-\frac{1}{2\sigma^2}\left(\sum_{i=1}^{n_1}(y_{1i}-\mu)^2+a_0\sum_{j=1}^{n_0}(y_{0j}-\mu)^2\right)\right\}\\
    &\sim \mathcal{N}\left(\frac{\sum_{i=1}^{n_1} y_{1i}+a_0\sum_{j=1}^{n_0} y_{0j}}{n_1+a_0n_0},\frac{\sigma^2}{n_1+a_0n_0}\right)_.
\end{align*}
Let $\tilde{Y}=\sum_{i=1}^{n_1} y_{1i}+a_0\sum_{j=1}^{n_0} y_{0j}$, $\tilde{n}=n_1+a_0n_0$, $\tilde{\mu}=\tilde{Y}/\tilde{n}$, and $\tilde{\sigma}^2=\sigma^2/\tilde{n}$. Consider the one-sided hypothesis $H_0: \mu\leq\mu_0$. The probability of rejecting $H_0$ is 
\begin{align*}
P(\mu>\mu_0|D_1,D_0,a_0)&=P\left(\frac{\mu-\tilde{\mu}}{\tilde{\sigma}}>\frac{\mu_0-\tilde{\mu}}{\tilde{\sigma}}\right)\\
&=\Phi\left( \frac{\tilde{\mu}-\mu_0}{\tilde{\sigma}} \right)_.
\end{align*}

At the $1-\alpha$ level, we reject $H_0$ when
\begin{align}\label{eq:rejectH0}
    &\Phi\left( \frac{\tilde{\mu}-\mu_0}{\tilde{\sigma}} \right)>1-\alpha \Longleftrightarrow
    \tilde{Y}>\tilde{n}(\mu_0+\Phi^{-1}(1-\alpha)\tilde{\sigma}).
\end{align}
To compute the type I error rate we assume that $H_0$ is true and furthermore that $\mu = \mu_0$, implying perfect compatibility, in which case $\text{E}(\tilde{Y})=(n_1+a_0n_0)\mu_0$ and $\text{Var}(\tilde{Y})=\sigma^2(n_1+a_0^2n_0)$. Then expression \eqref{eq:rejectH0} reduces to
\begin{align*}
    \frac{\tilde{Y}-\text{E}(\tilde{Y})}{\sqrt{\text{Var}(\tilde{Y})}}>\frac{\tilde{n}(\mu_0+\Phi^{-1}(1-\alpha)\tilde{\sigma})-\text{E}(\tilde{Y})}{\sqrt{\text{Var}(\tilde{Y})}}_,
\end{align*}
which occurs with probability
\begin{align}\label{eq:conservativeT1E}
    P\left(Z>\Phi^{-1}(1-\alpha)\sqrt{\frac{n_1+a_0n_0}{n_1+a_0^2n_0}}\right)_.
\end{align}

\begin{figure}[!htbp]
\begin{center}
\includegraphics[width=6in]{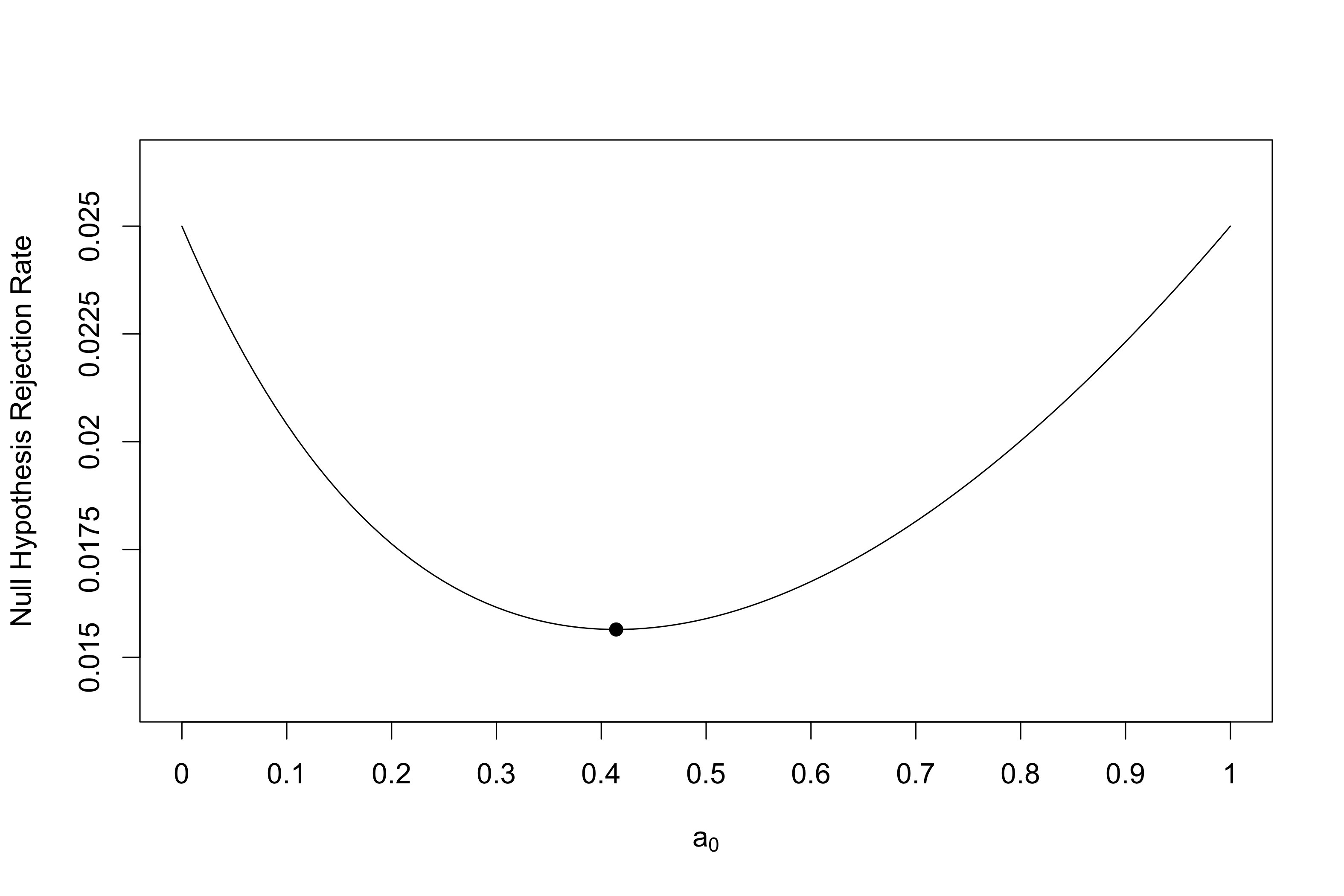}
\caption{Illustration of conservative type I error rate from equation \eqref{eq:conservativeT1E} when $n_1=n_0=100$.}
\label{fig:conservativeT1E}
\end{center}
\end{figure}

Note that expression \eqref{eq:conservativeT1E} is minimised when
\begin{align*}
    a_0=\sqrt{\frac{n_1(n_1+n_0)}{n_0^2}} - \frac{n_1}{n_0}_.
\end{align*}
For example, in the case that $n_1=n_0=100$ and $\alpha = 0.025$, the Type I Error is minimized when $a_0=\sqrt{2} - 1$ with a value of $0.01564$. This is reflected in Figure \ref{fig:conservativeT1E} where one can see that the type I error rate is less than 0.025 for every $a_0 \in (0,1)$. This result is consistent with the result observed for the fixed weight power prior in the perfect compatibility scenario.

\section*{Web Appendix D: Calibration Procedure}\label{calibration}

\subsection*{Case Weight Shrinkage Function}\label{sec:shrinkage}
\textcolor{black}{Web Appendix A empirically demonstrates in the setting of normally distributed endpoints that use of the untransformed case weighted power prior with compatible external data results in an inflated type I error rate due to producing biased estimates of the mean parameter. This inflated type I error rate was also observed in the setting of time-to-event endpoints using a proportional hazards regression model with piecewise constant baseline hazard. Web Appendix B proves in the setting of normally distributed endpoints that a fixed weight power prior with weight $a_0$ strictly between zero and one results in a conservative type I error rate, a property which was observed empirically in the survival setting. 
If there is perfect compatibility of the RCT controls and external controls, then the case weights will be uniformly distributed since the shared parameters are equivalent and the posterior predictive distribution is continuous \citep{Gelman2013}.
Based on these results, we transform the case weights by shrinking them towards the constant 0.5 (their expected value under perfect compatibility) to counterbalance the inflated type I error rate of the untransformed case weighted power prior with the conservative type I error rate of the fixed weight power prior to produce an analysis with controlled type I error rate at the nominal level.}

The level of shrinkage is determined via simulation such that the type I error rate is controlled at the nominal level for the given model (e.g. proportional hazards model with specific set of covariates) and given sample sizes for RWD and RCT data under the assumption of compatible external controls. Recall that $a_{j,k}$ is the case weight for interval $k$ of external control $j$. 

Consider the following the shifted and scaled polynomial function on the unit interval with $p\geq 1$:
\begin{align}\label{eq:cubic}
    f_p(a_{j,k}) = \frac{\text{sgn}(a_{j,k}-\frac{1}{2})|(2(a_{j,k} - 0.5))|^{p} + 1}{2}_,
\end{align}
where $\text{sgn}()$ is the sign function. This function will be used to shrink values of $a_{j,k}$ closer to $0.5$ while still allowing values of $a_{j,k}$ on the tails to approach zero and one (see Web Figure 2). The case weight shrinkage function from equation \eqref{eq:cubic} is defined for integer values of $p$ greater than or equal to one. Note that $p=1$ corresponds to the identity function on the unit interval, which would leave the case weights untransformed. As $p$ increases, the transformed case weights will converge to the fixed value of $0.5$, use of which leads to a conservative type I error rate as discussed in Web Appendix B. The value of $p$ is chosen as the smallest integer such that type I error is controlled for the specified design (see equation \eqref{eq:findp}). \textcolor{black}{Other types of shrinkage functions could be used and we do not claim the above family is optimal in any sense.}

\subsection*{Uniform Discounting Function}
We consider further modifying the case weights as described in Section \ref{sec:shrinkage} by applying a uniform discounting function to reflect data-set level incompatibility between external and RCT controls. %
Recall that $A=\{\mathbf{a}_1,...,\mathbf{a}_{n_0}\}$ is the collection of all subject- and interval-specific weights.
Define the average case weight for the external controls as
\begin{align}\label{eq:Abar}
    \overline{A}=\frac{\sum_{j=1}^{n_0}\sum_{k=1}^{K_j}a_{j,k}}{\sum_{j=1}^{n_0} K_j}.
\end{align}
Web Table 3 shows a family of logistic transformations based on the average case weight of the external controls based on the equation
\begin{align}\label{eq:logistic}
g_c(\overline{A}) = \frac{1}{1 + \exp(-q(\overline{A}-c))}_,
\end{align}

\textcolor{black}{where $c$ is a location parameter for the logistic transformation and $q$ is a specified shape parameter (e.g., $q = 50$). This class of transformations is applied to discount each case weight as the average case weight defined in equation \eqref{eq:Abar} begins to deviate from $0.5$, which is the expected value under perfect compatibility. Assigning a smaller location parameter $c$ results in greater discounting of the case weights for each average case weight value of $\overline{A}$. If the value of $\overline{A}$ is substantially less than $0.5$, then the discounting function will considerably discount all case weights towards zero. This results in an analysis that tends towards no borrowing, which attains the nominal type I error rate by construction.}

The value of $c$ will be chosen based on a predetermined level of maximum tolerated type I error rate in the case that there is a shift in baseline hazard for all external controls. A similar process was implemented by \cite{Psioda2018a}, and also could be framed as a predetermined maximum level of power reduction for a shift in baseline hazard for all external controls.

\subsection*{Combined Transformation Function}

 Combining the case weight shrinkage function from equation \eqref{eq:cubic} with the uniform discounting function from equation \eqref{eq:logistic}, the combined transformation function for a particular case weight is defined as
\begin{align}\label{eq:2part}
    h(a_{j,k},\overline{A})=f_p(a_{j,k})g_c(\overline{A}).
\end{align}

Using this two-part calibration procedure for the weights generalizes the weighted likelihood for the external controls with both subject- and interval-specific weights to 
\begin{align}\label{eq:f2-external-likeihood}
    \prod_{j=1}^{n_0}\mathcal{L}(\boldsymbol{\beta},\lambda|&\mathbf{D}_{0j}, h(\mathbf{a}_j,\overline{A}))) \nonumber \\ &=\prod_{j=1}^{n_0}\left\{(\lambda_{K_j}\text{exp}(x_j^T\boldsymbol{\beta}))^{h(a_{j,K_j},\overline{A})\nu_j} \prod_{k=1}^{K_j}\text{exp}\left\{ -h(a_{j,k},\overline{A})\lambda_k H_{j,k} \text{exp}(x_j^T\boldsymbol{\beta})\right\}\right\}_,
\end{align}
where $h(\mathbf{a}_j,\overline{A})=(h(a_{j,1},\overline{A}),...,h(a_{j,K_j},\overline{A}))$. 
This weighted likelihood for the external controls then will be used in the power prior from equation \eqref{eq:posterior_all}.

\textcolor{black}{In order to calibrate the type I error rate in the case of compatible external controls, consider the combined transformation function from equation \eqref{eq:2part} with $c=0$, which implies that $g_c(\overline{A})=1$ for any $\overline{A}>0$, and that $h(a_{j,k},\overline{A})$ reduces to $f_p(a_{j,k})$ from equation \eqref{eq:cubic}. This allows for the calibration of the type I error rate separate from the added uniform discounting. Let $H_0$ denote the hypothesis that the external controls are perfectly compatible with the RCT (i.e. baseline hazard and covariate parameter values are generated under the same mechanism as the RCT). To find the positive integer $p$ that results in controlled type I error, consider
\begin{align}\label{eq:findp}
    \underset{p\geq 1}{\text{argmin}} \hspace{0.05in} \left\lvert \text{Pr}\left( I \{\text{Pr}(\gamma_p > 0 | \mathbf{D}_1,\mathbf{D}_0) > 1 - \alpha \} | H_0\right) - \alpha\right\rvert,
\end{align}
where $\alpha$ is the nominal type I error rate (e.g. 0.025), and $\gamma_p$ is distributed according to the marginal posterior distribution of $\gamma$ from equation \eqref{eq:posterior_all} resulting from using the case weighted power prior based on the weighted likelihood in equation \eqref{eq:f2-external-likeihood} with $p$ defined in the case weight shrinkage function from equation \eqref{eq:cubic}.
}
\textcolor{black}{The value of $p$ from equation \eqref{eq:findp} is determined using a grid-search procedure over possible positive integers $p$ with external control data generated under $H_0$.}

\textcolor{black}{In order to calibrate the maximum type I error rate in the case of incompatible external controls with a shift in baseline hazard affecting all external controls, first consider the combined transformation function from equation \eqref{eq:2part} with $p$ from equation \eqref{eq:cubic} chosen from Step 1. It remains to find the value $c_{\alpha_{\text{max}}}$ that will be used in the uniform discounting function from equation \eqref{eq:logistic} that corresponds to the desired maximum type I error rate, denoted by $\alpha_{\text{max}}$. Let $H_{\text{shift}}$ denote the scenarios where all external controls have a shifted baseline hazard, but otherwise are compatible with the RCT model. To find the value of $c_{\alpha_{\text{max}}}$, consider
\begin{align}\label{eq:findc}
    \underset{c \in (0,0.5)}{\text{argmax}}:\hspace{0.025in}\text{Pr}\left(I\{\text{Pr}(\gamma_c > 0 | \mathbf{D}_1,\mathbf{D}_0) >1- \alpha\}|H_{\text{shift}}\right) < \alpha_{\text{max}},
\end{align}
where $\gamma_c$ is distributed according to the marginal posterior distribution of $\gamma$ from equation \eqref{eq:posterior_all} resulting from using the case weighted power prior based on the weighted likelihood in equation \eqref{eq:f2-external-likeihood} with $p$ defined in the case weight shrinkage function from equation \eqref{eq:cubic} and $c$ defined in the uniform discounting function from equation \eqref{eq:logistic}.
}
\textcolor{black}{The value of $c$ from equation \eqref{eq:findc} is determined using a grid-search procedure over possible values of $c\in(0,0.5)$ with external control data generated under $H_{\text{shift}}$.}

\section*{Web Appendix E: Case Weighted Commensurate Prior}
For the commensurate prior, the commensurability parameter is connected to the historical data likelihood with a link function, and the variance parameter of this link function governs the amount of borrowing. Let the hazard for $i^{th}$ patient be
\begin{align*}
    h_i(t|\boldsymbol{\lambda},\beta,\gamma)=h_0(t|\boldsymbol{\lambda})\text{exp}(z_i\gamma+x_i^T\beta+k_i\delta),
\end{align*}
where $k_i$ is the indicator that subject $i$ is an external control, and $\delta$ is the log drift hazard ratio between external and internal controls. We adopt the commensurate prior by assuming $\delta\sim\mathcal{N}(0,\sigma^2)$ and assuming a Cauchy hyperprior for variance of drift parameter representing commensurability. In particular, $\sigma\sim\text{Half-Cauchy}(0, 0.3)$.

The commensurate prior then has the form
    \begin{align}\label{eq:commensurate}
        \pi_0&(\gamma,\boldsymbol{\lambda},\boldsymbol{\beta},\delta,\sigma^2|D_0)\propto
        \left[\prod_{j=1}^{n_0}\mathcal{L}(\boldsymbol{\lambda},\boldsymbol{\beta},\delta,\sigma^2|D_{0j})\right]\pi_0(\delta|\sigma^2)\pi_0(\sigma^2)\pi_0(\gamma,\boldsymbol{\lambda},\boldsymbol{\beta}),
    \end{align}
    where $D_0=\{D_{01},...,D_{0n_0}\}$ with $D_{0j}$ being the data for $j^{th}$ external control, $\pi_0(\gamma,\boldsymbol{\lambda},\boldsymbol{\beta}$ is an initial prior, and $h_0(t|\lambda)$ is taken as a piecewise constant.
    
    The case weighted commensurate prior has the form
    \begin{align}\label{eq:weightedcommensurate}
        \pi_0&(\gamma,\boldsymbol{\lambda},\boldsymbol{\beta},\delta,\sigma^2|D_{0j})\propto
        \left[\prod_{j=1}^{n_0}\mathcal{L}(\boldsymbol{\beta},\boldsymbol{\lambda},\delta|\mathbf{D}_{0j}, \mathbf{a}_j) \right]\pi_0(\delta|\sigma^2)\pi_0(\sigma^2)\pi_0(\gamma,\boldsymbol{\lambda},\boldsymbol{\beta}),
    \end{align}
where 
\begin{align*}
    \mathcal{L}(\boldsymbol{\beta},\boldsymbol{\lambda},\delta|\mathbf{D}_0, \mathbf{a}_j)  =(\lambda_{K_j}\text{exp}(x_j^T\boldsymbol{\beta} + k_i\delta))^{a_{j,K_j}\nu_j}\prod_{k=1}^{K_j}\text{exp}\left\{ -a_{j,k}\lambda_k H_{j,k} \text{exp}(x_j^T\boldsymbol{\beta} + k_i\delta)\right\}.
\end{align*}

\section*{Web Appendix F: Additional Simulation Study Results}

\subsection*{Partial Shift Confounding}

\begin{figure}[!htbp]
\begin{center}
\includegraphics[width=6.25in]{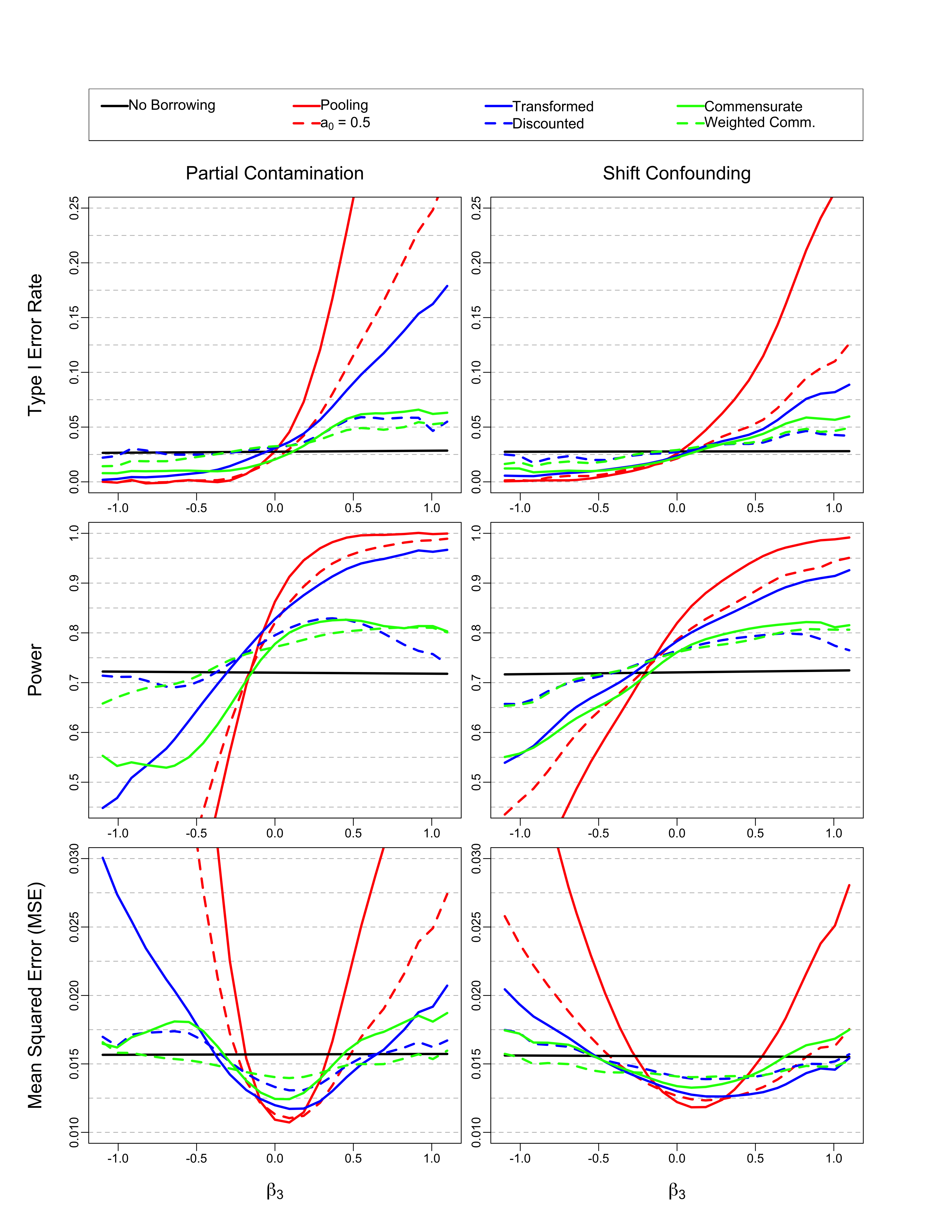}
\caption{Partial shift confounding scenario; selected operating characteristics by analysis method. Transformed = transformed case weighted power prior; Discounted = discounted transformed case weighted power prior with maximum type I error rate under shift confounding calibrated at 0.15.}
\label{fig:6panel-partial-shift}
\end{center}
\end{figure}

\subsection*{Vary proportion of external controls}
\begin{figure}[!htbp]
\begin{center}
\includegraphics[width=6.25in]{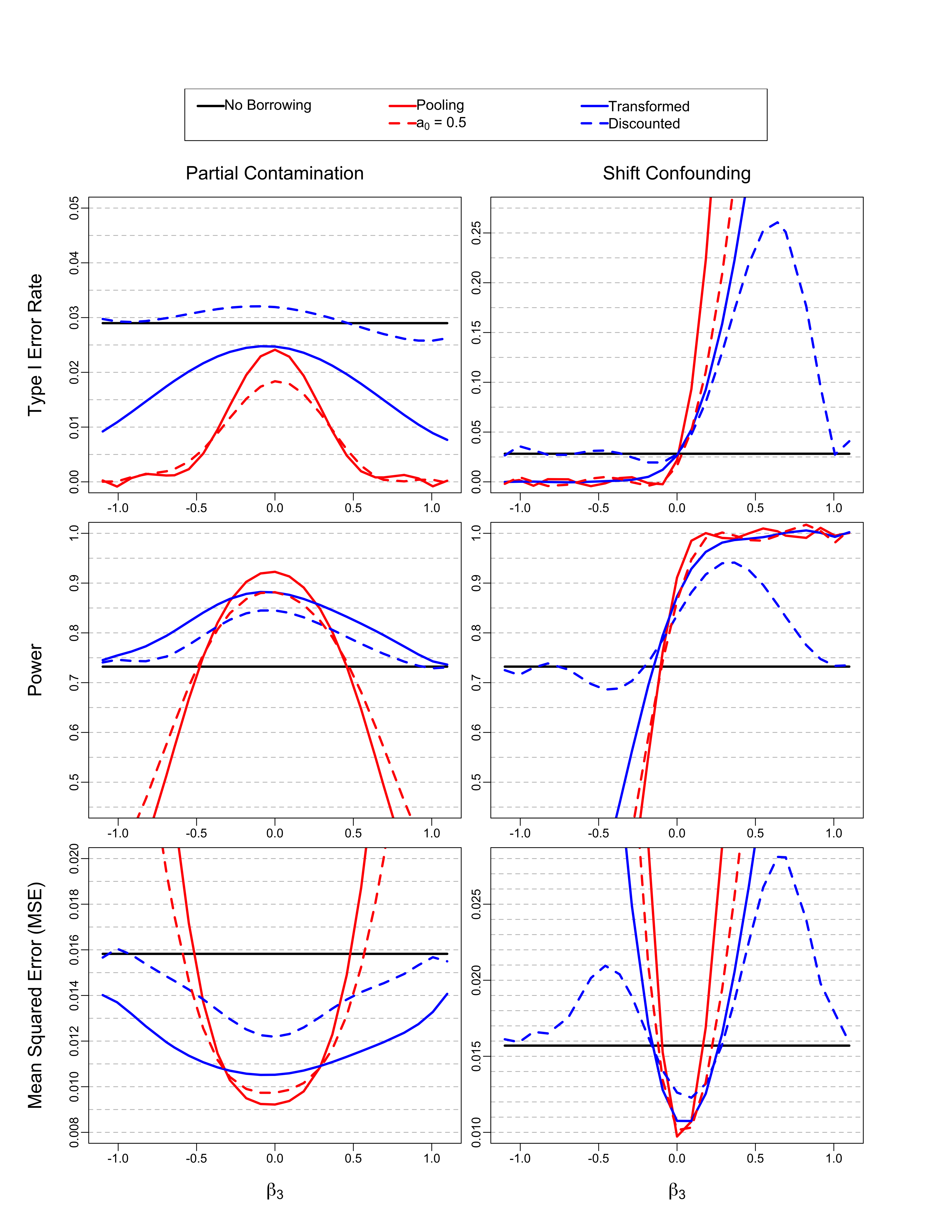}
\caption{Increasing external controls to 200; selected operating characteristics by analysis method. Transformed = transformed case weighted power prior; Discounted = discounted transformed case weighted power prior with maximum type I error rate under shift confounding calibrated at 0.15.}
\label{fig:6panel-partial-shift}
\end{center}
\end{figure}
\begin{figure}[!htbp]
\begin{center}
\includegraphics[width=6.25in]{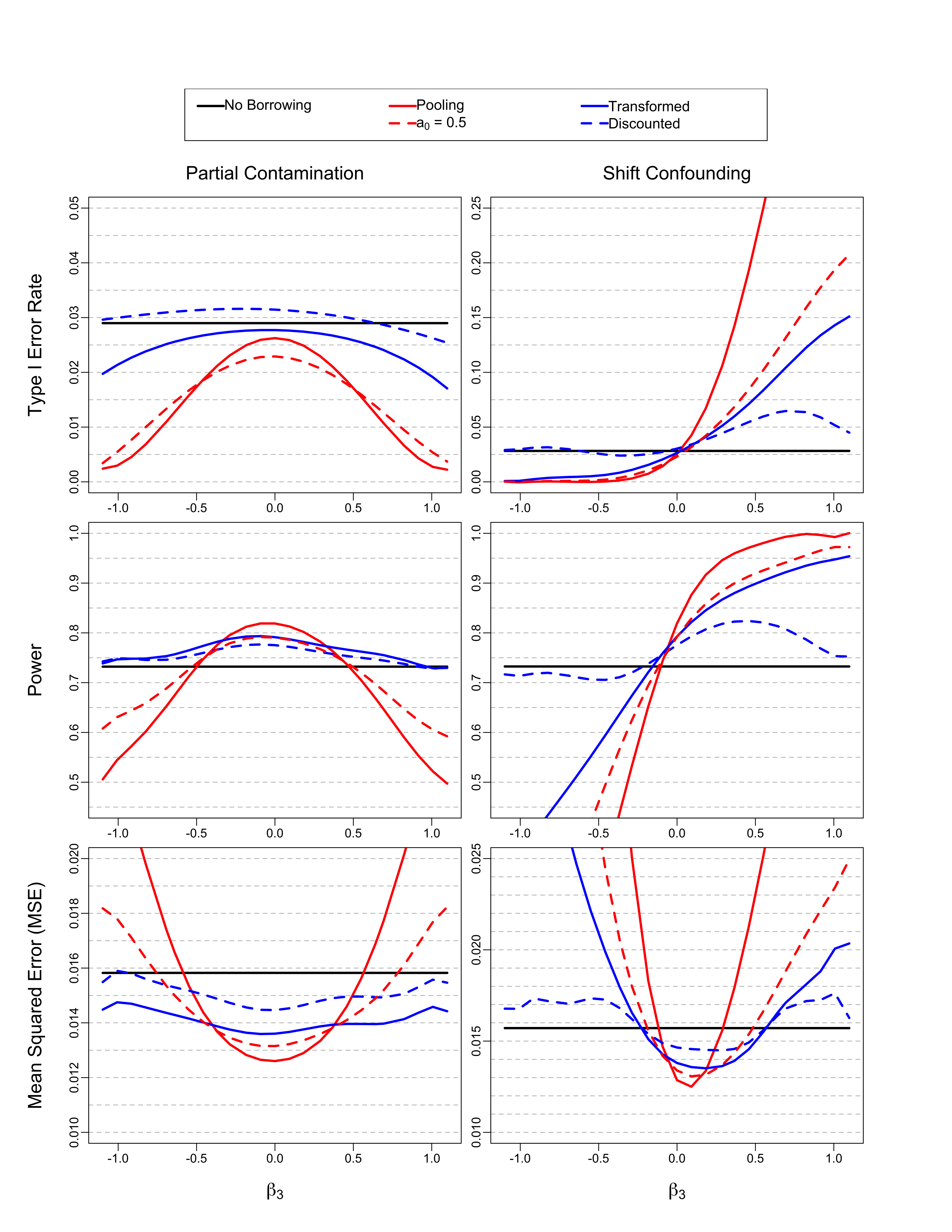}
\caption{Decreasing external controls to 50; selected operating characteristics by analysis method. Transformed = transformed case weighted power prior; Discounted = discounted transformed case weighted power prior with maximum type I error rate under shift confounding calibrated at 0.15.}
\label{fig:6panel-partial-shift}
\end{center}
\end{figure}
\subsection*{Study impact of model misspecification}
\begin{figure}[!htbp]
\begin{center}
\includegraphics[width=6.25in]{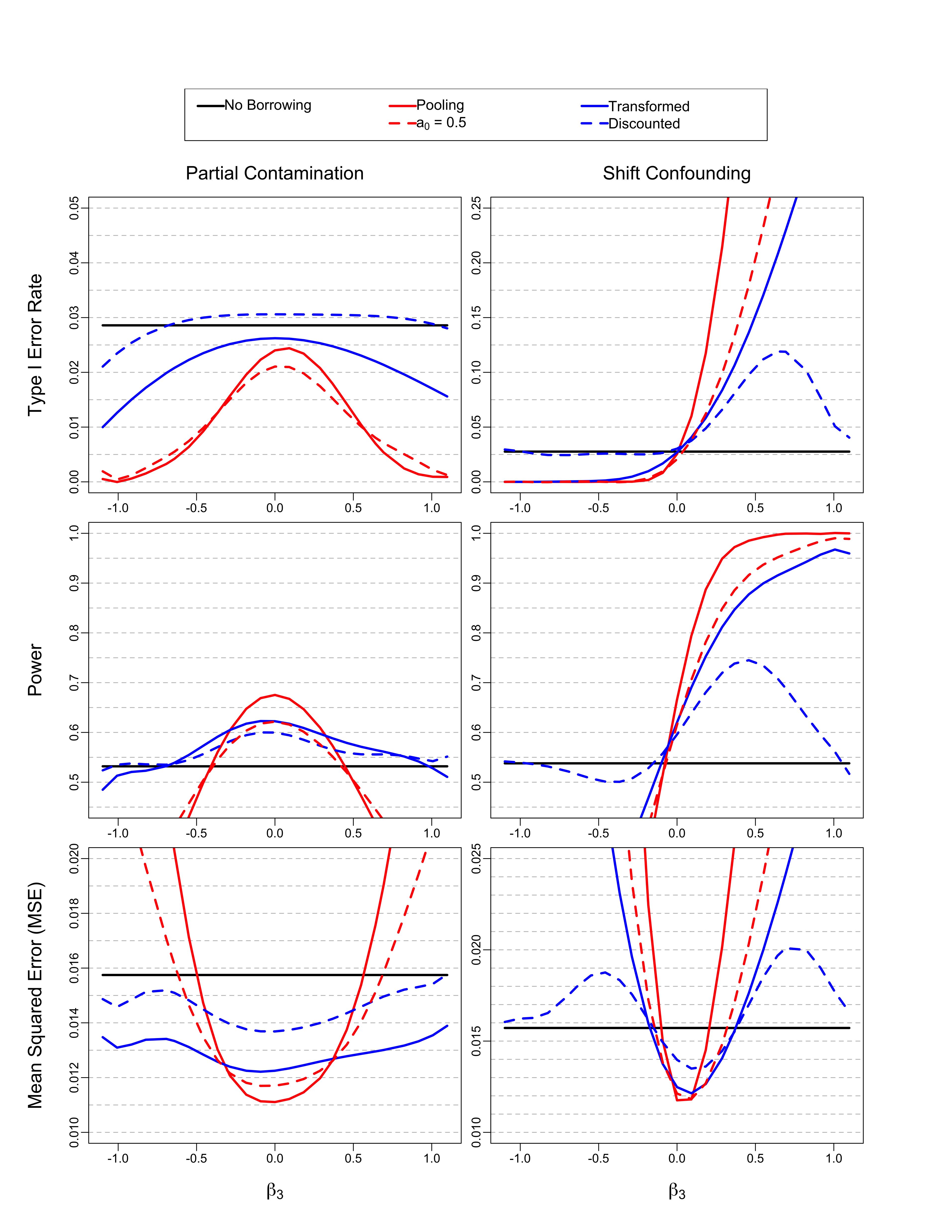}
\caption{Delayed treatment effect through 100 days; selected operating characteristics by analysis method. Transformed = transformed case weighted power prior; Discounted = discounted transformed case weighted power prior with maximum type I error rate under shift confounding calibrated at 0.15.}
\label{fig:6panel-partial-shift}
\end{center}
\end{figure}
\begin{figure}[!htbp]
\begin{center}
\includegraphics[width=6.25in]{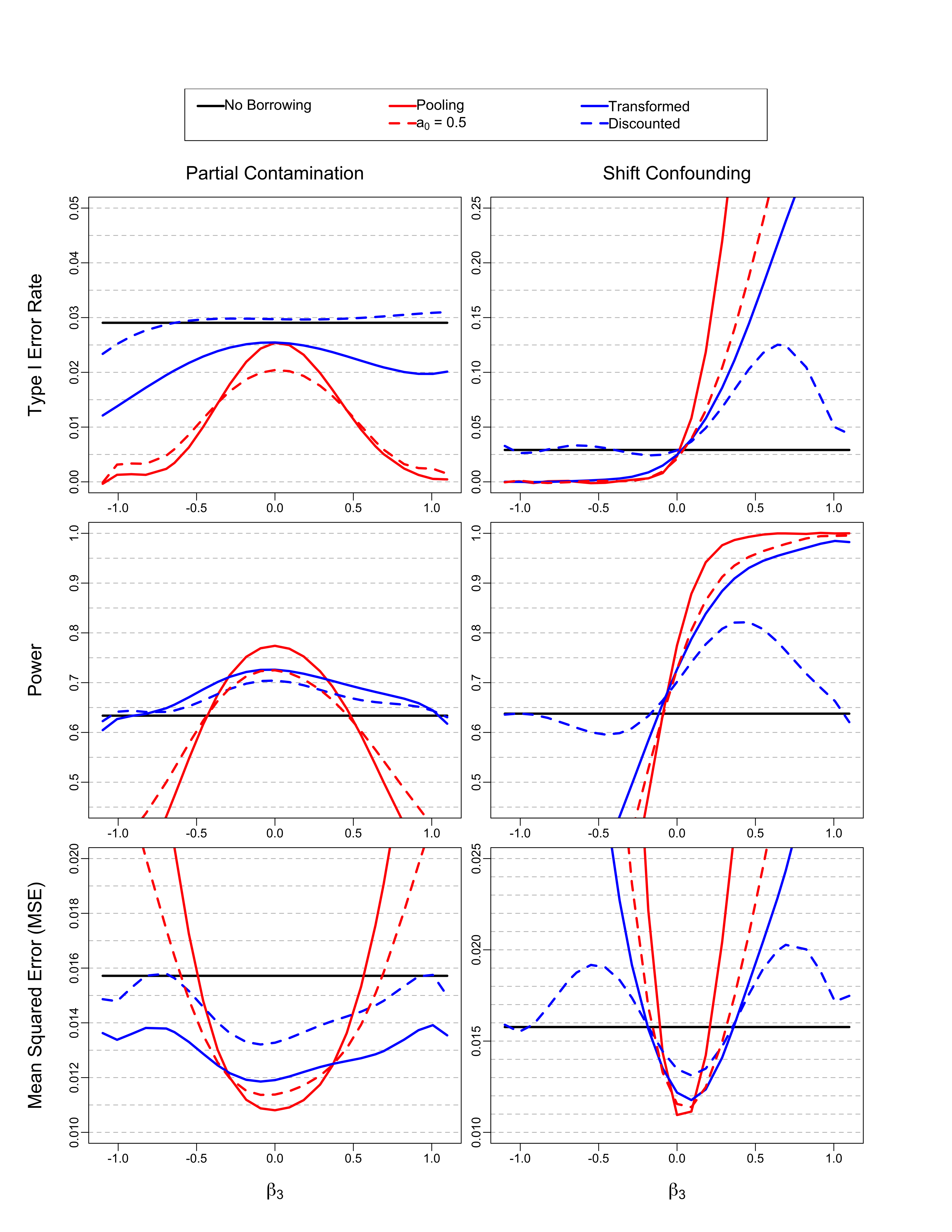}
\caption{Delayed treatment effect through 50 days; selected operating characteristics by analysis method. Transformed = transformed case weighted power prior; Discounted = discounted transformed case weighted power prior with maximum type I error rate under shift confounding calibrated at 0.15.}
\label{fig:6panel-partial-shift}
\end{center}
\end{figure}

\subsection*{Choosing Number of Segments for Baseline Hazard}
\begin{table}[p]
\centering
\caption{Treatment effect estimation and model fit diagnostics single generated datasets under different magnitudes of shift confounding. $K_G$: number of baseline hazard segments used in the data generating mechanism; $K_M$: number of baseline hazard segments used in the analysis model. }
\begin{tabular}{ccccccccccccc}
  \hline
& & & & 
\multicolumn{3}{c}{Adaptive} & 
\multicolumn{3}{c}{No Borrowing} &
\multicolumn{3}{c}{Pooling}\\
                            \cline{5-13}
$\beta_3$	&	$K_G$	&	$K_M$	&	$\overline{A}$	&	HR	&	CI W.	&	BIC	&	HR	&	CI W.	&	BIC	&	HR	&	CI W.	&	BIC	\\	
   \hline
	&		&	1	&	0.525	&	0.737	&	0.319	&	5754.9	&	0.718	&	0.352	&	5756.2	&	0.773	&	0.314	&	5753.8	\\	
0	&	1	&	3	&	0.484	&	0.684	&	0.299	&	5764.6	&	0.703	&	0.347	&	5764.1	&	0.757	&	0.310	&	5760.9	\\	
	&		&	5	&	0.510	&	0.696	&	0.306	&	5775.0	&	0.704	&	0.347	&	5775.7	&	0.757	&	0.310	&	5772.4	\\	
   \hline
	&	&	1	&	0.315	&	0.818	&	0.371	&	6123.5	&	0.718	&	0.352	&	6181.7	&	1.558	&	0.642	&	6005.1	\\	
-1.1	&	1	&	3	&	0.435	&	0.882	&	0.393	&	6125.1	&	0.703	&	0.347	&	6230.1	&	1.551	&	0.650	&	6014.3	\\	
	& &	5	&	0.466	&	0.931	&	0.420	&	6122.8	&	0.704	&	0.347	&	6240.4	&	1.550	&	0.650	&	6025.6	\\	
   \hline
	&		&	1	&	0.385	&	0.632	&	0.280	&	5604.2	&	0.718	&	0.352	&	5616.7	&	0.506	&	0.204	&	5594.8	\\	
1.1	&	1	&	3	&	0.464	&	0.608	&	0.270	&	5608.5	&	0.703	&	0.347	&	5626.4	&	0.509	&	0.210	&	5600.1	\\	
	&		&	5	&	0.518	&	0.595	&	0.264	&	5619.2	&	0.704	&	0.347	&	5638.7	&	0.509	&	0.210	&	5612.0	\\	
   \hline
	&		&	1	&	0.292	&	0.690	&	0.326	&	7513.7	&	0.639	&	0.316	&	7511.6	&	0.610	&	0.258	&	7511.1	\\	
0	&	3	&	3	&	0.554	&	0.736	&	0.327	&	6922.3	&	0.716	&	0.352	&	6924.0	&	0.768	&	0.326	&	6920.8	\\	
	&		&	5	&	0.548	&	0.754	&	0.337	&	6934.1	&	0.717	&	0.353	&	6941.9	&	0.769	&	0.327	&	6932.3	\\	
   \hline
\end{tabular}
\label{tbl:pick_nseg}
\end{table}

\end{document}